\documentclass[prl,twocolumn,floatfix,superscriptaddress,amsmath,citeautoscript,aps,longbibliography]{revtex4-2}
\pdfoutput=1
\usepackage[T1]{fontenc}\usepackage[latin1]{inputenc}
\usepackage{dcolumn,bm,csquotes,graphicx,color,booktabs,microtype,afterpage} \graphicspath{{Figures/}}
\usepackage[charter,greekuppercase=italicized]{mathdesign}
\renewcommand{\figurename}{Fig.}
\renewcommand{\tablename}{Table}
\makeatletter\renewcommand{\fnum@figure}[1]{\figurename~\thefigure.}\makeatother
\makeatletter\renewcommand{\fnum@table}[1]{\tablename~\thetable.}\makeatother
\newcount\hh \newcount\mm
\hh=\time \divide\hh by 60
\mm=\hh \multiply\mm by 60 \mm=-\mm
\advance\mm by \time
\def\now{\number\hh:\ifnum\mm<10{}0\fi\number\mm}

\DeclareMathAlphabet{\cmcal}{OMS}{cmsy}{m}{n}

\usepackage{multibib}
\newcites{S}{Supplementary References}

\usepackage[colorlinks,plainpages=false,linkcolor=blue,urlcolor=blue,citecolor=blue,pdfpagemode=UseNone,pdfstartview=FitBH]{hyperref}
\makeatletter\AddToHook{env/thebibliography/end}{\bibitem@fin\let\@bibstop\@undefined\edef\@currentlabel{\arabic{NAT@ctr}}\label{LastBibItem}}\makeatother
\AtBeginDocument{}

\newcommand{\zncr}{ZnCr$_2$Se$_4$}

\usepackage{soul}

\begin{document}
\title{Instability of single-${\it q}$ helimagnets as an indicator of non-Heisenberg interactions}

\author{A.\,Sanders}
\affiliation{Helmholtz-Zentrum Berlin f{\"u}r Materialien und Energie, Hahn-Meitner-Platz 1, 14109 Berlin, Germany}
\affiliation{Dahlem Center for Complex Quantum Systems and Fachbereich Physik, Freie Universit{\"a}t Berlin, Arnimallee 14, 14195 Berlin, Germany}
\author{N.\,D.\,Andriushin}
\affiliation{Institut f{\"u}r Festk{\"o}rper- und Materialphysik, Technische Universit{\"a}t Dresden, D-01069 Dresden, Germany}
\author{D.\,C.\,Peets}
\affiliation{Institut f{\"u}r Festk{\"o}rper- und Materialphysik, Technische Universit{\"a}t Dresden, D-01069 Dresden, Germany}
\author{P.~Steffens}
\affiliation{Institut Laue-Langevin, 71 avenue des Martyrs, CS 20156, 38042 Grenoble Cedex 9, France}
\author{A.~Hiess}
\affiliation{Institut Laue-Langevin, 71 avenue des Martyrs, CS 20156, 38042 Grenoble Cedex 9, France}
\affiliation{European Spallation Source ERIC, P.O. Box 176, 221 00 Lund, Sweden}
\author{L.\,Prodan}
\affiliation{Experimentalphysik V, Center for Electronic Correlations and Magnetism, Institute of Physics, University of Augsburg, D-86159 Augsburg, Germany}
\author{V.\,Tsurkan}
\affiliation{Experimentalphysik V, Center for Electronic Correlations and Magnetism, Institute of Physics, University of Augsburg, D-86159 Augsburg, Germany}
\affiliation{Institute of Applied Physics, Moldova State University, MD 2028 Chisinau, Republic of Moldova}
\author{V.\,Felea}
\affiliation{Institute of Applied Physics, Moldova State University, MD 2028 Chisinau, Republic of Moldova}
\author{D.\,S.\,Inosov}
\affiliation{Institut f{\"u}r Festk{\"o}rper- und Materialphysik, Technische Universit{\"a}t Dresden, D-01069 Dresden, Germany}
\affiliation{W\"urzburg-Dresden Cluster of Excellence on Complexity and Topology in Quantum Matter\,---\,\textit{ct.qmat}, TU Dresden}
\author{J.\,Reuther}
\affiliation{Helmholtz-Zentrum Berlin f{\"u}r Materialien und Energie, Hahn-Meitner-Platz 1, 14109 Berlin, Germany}
\affiliation{Dahlem Center for Complex Quantum Systems and Fachbereich Physik, Freie Universit{\"a}t Berlin, Arnimallee 14, 14195 Berlin, Germany}

\begin{abstract}\noindent

Single-$\bm{q}$ planar spirals in frustrated Heisenberg magnets possess anomalously soft deformations whose energy grows only quartically in their amplitude. We investigate under which conditions additional interactions generate harmonic mass terms for these quartic modes and thereby destabilize the spiral towards multi-$\bm{q}$ order\,---\,a Landau-type competition that is realized already on the level of the Hamiltonian. For helimagnets on the pyrochlore lattice we show that no bilinear anisotropic interaction compatible with the lattice symmetries can produce such an instability, whereas a small isotropic biquadratic interaction does. Elastic neutron scattering on the helimagnet ZnCr$_2$Se$_4$ reveals a weak magnetic signal at pseudo-Goldstone wave vectors, indicating multi-$\bm{q}$ order, which we reproduce in classical Monte Carlo simulations including a biquadratic coupling. An instability of single-$\bm{q}$ helimagnetic order can thus serve as a sensitive indicator of spin interactions beyond the bilinear level.
\end{abstract}

\maketitle

\begin{figure}
  \includegraphics[width=0.8\columnwidth]{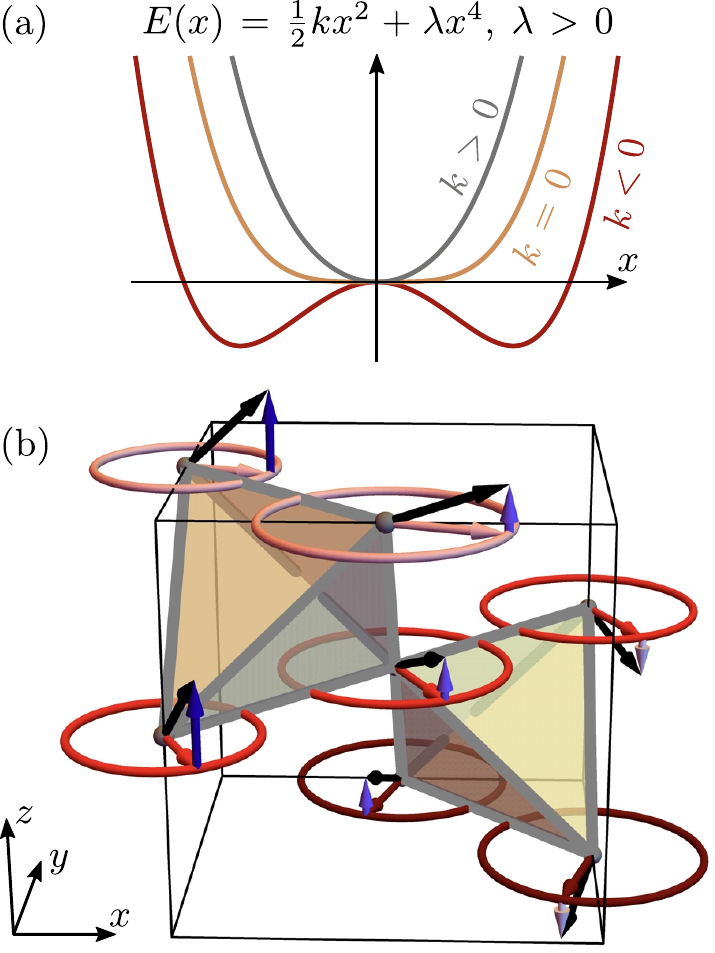}
	\caption{ (a) Sum of quadratic and quartic potentials, where $k<0$ gives rise to minima at $x\neq0$. (b) Deformed spin state of Eq.~(\ref{eq:quartic_deformation}) (black arrows) on the pyrochlore lattice. The spiral configuration of the unperturbed planar state has wave vector ${\bm q}_\text{B}=(0,0,\xi)$, propagating along the vertical direction, as indicated by the three shades of red used for the in-plane arrows. The quartic deformation introduces an additional out-of-plane component with wave vector ${\bm q}_\text{PG}=(\xi,0,0)$, propagating along the $x$ axis, as indicated by the three shades of blue used for the out-of-plane arrows.}
	\label{fig:spin_state}
\end{figure}

\textit{Introduction.}--- Landau theory is one of the cornerstones of condensed matter physics. It provides a unified description of phases of matter and of the transitions between them in terms of spontaneous symmetry breaking~\cite{Landau1936,Landau1937a,Landau1937b}. Its central object is the Landau potential $F(x)$ of an order parameter $x$, which in the simplest case is governed by the competition between a quadratic (harmonic or mass) term $\sim\!x^2$ and a quartic term $\sim\!x^4$. The sign of the quadratic term decides whether the phase with $\langle x\rangle=0$ is stable (positive mass) or unstable (negative mass), see \autoref{fig:spin_state}(a). In the latter case the system develops $\langle x\rangle\neq0$, which is typically accompanied by the breaking of a symmetry. Usually, the sign of the mass term is controlled by the strength of thermal or quantum fluctuations.

In this Letter, we study a less common situation: a classical many-body system in which the competition between quadratic and quartic terms is already present at the level of the Hamiltonian, i.e., at $T=0$ and in the absence of any fluctuations. Whether the system realizes $\langle x\rangle=0$ or $\langle x\rangle\neq0$ then carries no information about fluctuations, but instead reveals properties of the Hamiltonian itself. We demonstrate this simple principle theoretically and experimentally for a frustrated spin system and show that it can be surprisingly rich and sensitive even to small terms in the Hamiltonian.

Our starting point is a classical Heisenberg model with competing interactions whose ground states are single-$\bm q$ planar spirals,
\begin{equation}
{\bm S}_{i\alpha}^0=\left(\cos(\bm q_n \cdot\bm r_{i\alpha}+\phi),\sin(\bm q_n \cdot\bm r_{i\alpha}+\phi),0\right),\label{eq:planar_spiral}
\end{equation}
where, without loss of generality, we have chosen the $x$-$y$ plane as the spiral plane. Here, $\bm q_n$ is a ground state wave vector, $\phi$ is an arbitrary phase, and $\bm r_{i\alpha}$ is the position of the site in unit cell $i$ and sublattice $\alpha$. While our considerations can be applied to arbitrary inversion-symmetric lattices, our main focus is on the pyrochlore lattice and an application to \zncr. Crystal symmetries usually imply a set of symmetry-related ground-state wave vectors $\bm q_n$, labeled by $n$, one of which, referred to as the Bragg wave vector $\bm q_\text{B}$, is spontaneously selected at low temperatures. In this situation the spiral admits a deformation, parametrized by $x$, of the form
\begin{equation}\label{eq:quartic_deformation}
{\bm S}_{i\alpha}=\frac{1}{N_{i\alpha}}\left[\bm S_{i\alpha}^0+x\left(0,0,\cos(\bm q_\text{PG}\cdot \bm r_{i\alpha}+\phi')\right)\right],
\end{equation}
whose energy grows only {\it quartically} in $x$, i.e., $E(x)=\lambda x^4+\mathcal{O}(x^6)$ with a system-dependent coefficient $\lambda>0$. The existence of this quartic deformation was established in Ref.~\cite{Fancelli2026} for Bravais lattices and is generalized here to non-Bravais lattices. In Eq.~(\ref{eq:quartic_deformation}), $\bm q_\text{PG}$ is a so-called pseudo-Goldstone wave vector, which is one of the symmetry-related ground-state wave vectors $\bm q_n$ that is {\it not} spontaneously selected, $\bm q_\text{PG}\neq\bm q_\text{B}$. Furthermore, $\phi'$ is another arbitrary phase and $N_{i\alpha}$ is a site-dependent normalization factor to ensure $|{\bm S}_{i\alpha}|=1$. It is essential that the deformation in Eq.~(\ref{eq:quartic_deformation}) is perpendicular to the spiral plane, see \autoref{fig:spin_state}(b) for an illustration. This situation is prominently realized in the helimagnet \zncr\ which features a pyrochlore lattice of spin-$3/2$ Cr$^{3+}$ ions and ordering wave vectors $\bm q_n=(\xi,0,0)$, $(0,\xi,0)$, $(0,0,\xi)$ with $\xi=0.44\text{ r.l.u.}$~\cite{Cameron2016,Tymoshenko_2017}. As demonstrated by inelastic neutron scattering in our recent companion paper~\cite{Andriushin2026}, the quartic deformation gives rise to a thermally enhanced pseudo-Goldstone mode: an anomalously soft excitation whose gap {\it increases} with temperature~\cite{Fancelli2026,Khatua2023} (see also Refs.~\cite{Rau2018,Hickey2025} for pseudo-Goldstone modes from quantum fluctuations).

The existence of this quartic potential immediately raises the question of whether non-Heisenberg interactions can destabilize the single-$\bm q$ planar spiral in Eq.~(\ref{eq:planar_spiral}) by generating a harmonic term, such that $E(x)=\frac{1}{2}k x^2+\lambda x^4$. For $k<0$, the deformation in Eq.~(\ref{eq:quartic_deformation}) acquires a finite amplitude in the ground state, which corresponds to an instability towards a multi-$\bm q$ state. 

At first sight, one might expect such mass terms to arise naturally from perturbations in the form of spin-anisotropic interactions, for which the mechanism of Ref.~\cite{Fancelli2026} no longer applies. We show, however, that the single-$\bm q$ spiral is surprisingly robust. Specifically, for a helimagnet, i.e., when the normal of the spiral plane is parallel to $\bm q_\text{B}$, no bilinear spin-anisotropic interaction compatible with the crystal symmetries generates a nonzero mass term.

This robustness creates a puzzle for \zncr, for which we report elastic neutron scattering data indicating multi-$\bm q$ magnetic order, with a dominant signal at the magnetic Bragg wave vector $\bm q_\text{B}$ and a weaker signal at the pseudo-Goldstone wave vectors $\bm {q}_{\text{PG}1},\bm{q}_{\text{PG}2}$. After ruling out other possible sources of this signal, we argue that a small isotropic biquadratic spin interaction is responsible for the instability towards a \textit{double}-$\bm q$ spiral, a scenario that we confirm by Monte Carlo simulations. An observed multi-$\bm q$ helimagnetic order can therefore serve as a highly sensitive indicator of a small biquadratic spin interaction.

\textit{Soft deformations of planar spin spirals.}---
We first briefly revisit the formation of soft modes in classical spin systems with planar spiral ground states, generalizing the analysis of Ref.~\cite{Fancelli2026} for the square lattice to arbitrary inversion-symmetric non-Bravais lattices. Our starting point is a general classical Heisenberg Hamiltonian
\begin{equation}
H=\frac{1}{2}\sum_{ij}\sum_{\alpha\beta} J_{ij\alpha\beta}{\bm S}_{i\alpha}\cdot {\bm S}_{j\beta}.\label{eq:ham}
\end{equation}
As explained in detail in the Supplemental Material (SM)~\cite{SM}, if Eq.~(\ref{eq:ham})
is Luttinger-Tisza solvable~\cite{Luttinger1946,Luttinger1951}, and if the lowest eigenvector of the Fourier-transformed coupling matrix $J_{\alpha\beta}(\bm k=\bm q_n)$ at each ground-state wavevector $\bm q_n$ has equal
entries on all sublattices $\alpha$, then the spirals of Eq.~(\ref{eq:planar_spiral}) are exact classical ground states. 

We suppose that the particular ground state with Bragg propagation vector $\bm{q}_{\rm B}$ has been spontaneously chosen, and work in local, rotating coordinate frames such that the spiral of Eq.~(\ref{eq:planar_spiral}) takes the form ${\bm S}_{i\alpha}^\text{loc}=(1,0,0)$. Arbitrary deformations can then be parametrized as
\begin{equation}\label{eq:general_deformation}
{\bm S}_{i\alpha}^\text{loc}=(1,\Delta^\parallel_{i\alpha},\Delta^\perp_{i\alpha})/\sqrt{1+(\Delta^\parallel_{i\alpha})^2+(\Delta^\perp_{i\alpha})^2},
\end{equation}
where $\Delta^\parallel_{i\alpha}$ ($\Delta^\perp_{i\alpha}$) describes an in-plane (out-of-plane) deformation. The corresponding normal modes and their harmonic energies follow from inserting Eq.~(\ref{eq:general_deformation}) into Eq.~(\ref{eq:ham}) and diagonalizing the Hessian $\partial^2 H/\partial \Delta_{i\alpha}\partial \Delta_{j\beta}$ for both types of deformations, $\Delta_{i\alpha}\in\{\Delta^\parallel_{i\alpha},\Delta^\perp_{i\alpha}\}$~\cite{Walker1980,Noculak2023,Lozano2024}. As shown in the SM~\cite{SM}, this can be done analytically in Fourier space, yielding the harmonic-order zero modes
\begin{equation}\label{eq:delta}
\Delta_{i\alpha}^\perp=\sum_{n} \left[a_{\bm{q}_n}\cos(\bm q_n\cdot \bm r_{i\alpha})+b_{\bm{q}_n}\sin(\bm q_n\cdot \bm r_{i\alpha})\right],\quad \Delta_{i\alpha}^\parallel=c_{\bm{q}_{\rm B}}.
\end{equation}
Here, $n$ runs over all physically distinct ground-state wavevectors, $n\in\{\rm B, PG1, PG2,\dots\}$, comprising the Bragg wave vector ${\bm q}_\text{B}$ and the remaining pseudo-Goldstone wave vectors $\bm q_{\text{PG}n},\,n=1,2,\dots$. 
The free parameters $a_{\bm q_n}$, $b_{\bm q_n}$, and $c_{\bm{q}_{\rm B}}$ therefore span the space of symmetry-enforced, quadratic-order zero modes. 

The homogeneous in-plane deformation $\Delta_{i\alpha}^\parallel=c_{\bm{q}_{\rm B}}$ is a Goldstone mode corresponding to global spin rotations about the $z$ axis. Since a planar spiral breaks all continuous spin-rotation symmetries, two further Goldstone modes exist at $\bm q_n=\bm q_\text{B}$, parametrized by $a_{\bm q_\text{B}}$ and $b_{\bm q_\text{B}}$; they correspond to global rotations about in-plane axes. Because they originate from symmetries, the energy of these Goldstone modes vanishes to all orders. In contrast, the out-of-plane deformations $a_{\bm q_{\text{PG} n}}$ and $b_{\bm q_{\text{PG} n}}$ are `accidental' harmonic zero modes, and so generically lie in a \textit{quartic} restoring potential.

The cos and sin components at a given $\bm q_\text{PG}$ in Eq.~(\ref{eq:delta}) are related by a phase shift, i.e., a translation along $\bm q_\text{PG}$. Up to this distinction, there is one quartic mode for each pair of pseudo-Goldstone wave vectors $\pm \bm q_\text{PG}$. For \zncr, where the $\bm q_{n}$ lie along the three cubic axes, this amounts to two quartic modes at $\bm q_{\text{PG1}}$ and $\bm q_{\text{PG2}}$.

\textit{Bilinear anisotropic perturbations.}---
The formation of quartic modes relies on the Heisenberg form of the interactions. One might therefore expect that general bilinear anisotropic perturbations $V$ to $H$ generate additional quadratic potentials. This would realize the Landau scenario of \autoref{fig:spin_state}(a), in which the single-$\bm q$ state can become unstable towards a multi-$\bm q$ state. To investigate this possibility, we consider a general bilinear anisotropic perturbation
\begin{equation}
V=\frac{1}{2}\sum_{ij}\sum_{\alpha\beta}\sum_{\mu\nu}S^\mu_{i\alpha} V^{\mu\nu}_{ij\alpha\beta} S^\nu_{j\beta},\label{eq:ham_aniso}
\end{equation}
where $\mu,\nu=x,y,z$ label spin components. We expand $V$ up to quadratic order in the space of quartic modes, $V\approx V_0+\bm m\cdot \bm \delta+\frac{1}{2}\bm \delta^\text{T} M \bm\delta$, where $\bm \delta =(a_{\bm q_{\text{PG1}}},b_{\bm q_{\text{PG1}}},a_{\bm q_{\text{PG2}}},b_{\bm q_{\text{PG2}}})$. Here, $V_0$ is the energy of the planar spiral $\bm S^0_{i\alpha}$ due to the anisotropic perturbation, $\bm m$ is a linear drift term of the quartic-mode amplitude, and the eigenvalues of $M$ are the mass terms acquired by the pseudo-Goldstone modes. For translation-invariant $V$, $\bm m$ vanishes and the quartic modes do not couple to the Goldstone modes in harmonic order.

One subtlety deserves a comment. Deformation modes of $H$ at wave vectors close to a given $\bm q_n$ have a small mass, which competes directly with the quadratic contribution of $V$ at the same wave vectors and may change the sign of the total mass in $H+V$. As a consequence, the wave vectors $\bm q_n$ may be modified by $V$. In the generic case they are merely rescaled by a constant $\zeta$, i.e., $\bm q_n\rightarrow \zeta\bm q_n$. Our analysis of $V$ in the space of quartic modes then still applies, albeit at the renormalized wave vectors.

We define $V_{\alpha\beta}^{\mu\nu}(\bm k)$ as the Fourier transform of $V^{\mu\nu}_{ij\alpha\beta}$ and $V^{\mu\nu}(\bm k)=\sum_{\alpha\beta}V_{\alpha\beta}^{\mu\nu}(\bm k)$, which is real for inversion-symmetric lattices. Taking the $xy$ plane as the spiral plane, we find~\cite{SM} that $M$ is diagonal,
\begin{align}
M=&-\frac{N_u}{4}\left[V^{xx}(\bm q_\text{B})+V^{yy}(\bm q_\text{B})\right]\text{diag}(1,1,1,1)\notag\\
&+\frac{N_u}{2}\,\text{diag}\left(V^{zz}(\bm q_{\text{PG1}}),V^{zz}(\bm q_{\text{PG1}}),V^{zz}(\bm q_{\text{PG2}}),V^{zz}(\bm q_{\text{PG2}})\right),\label{eq:masses}
\end{align}
where $N_u$ is the number of unit cells.

We now apply this result to \zncr, whose helimagnetic order implies $\bm q_\text{B}=(0,0,\xi)$, $\bm q_\text{PG1}=(\xi,0,0)$, and $\bm q_\text{PG2}=(0,\xi,0)$. The mirror symmetry of the pyrochlore lattice that interchanges $x$ and $y$ yields $V^{xx}(\bm q_\text{B})=V^{yy}(\bm q_\text{B})$ and $V^{zz}(\bm q_{\text{PG1}})=V^{zz}(\bm q_{\text{PG2}})$. Similarly, the mirror symmetry interchanging $x$ and $z$ gives $V^{zz}(\bm q_{\text{PG1}})=V^{xx}(\bm q_\text{B})$, such that $M$ in Eq.~(\ref{eq:masses}) vanishes identically. Surprisingly, this means that as long as the pyrochlore lattice symmetries are intact, no bilinear anisotropic spin interaction can turn the quartic modes of a helimagnet into quadratic ones. As we show in the SM~\cite{SM}, however, the quartic modes may become unstable if the helimagnetic condition is relaxed, i.e., when the spiral acquires a cycloidal component.

\textit{Multi-$\bm q$ order in \zncr\ from neutron scattering.}---
We next report elastic neutron scattering data (see End Matter for experimental methods) on the magnetic order of \zncr\ and then reconcile them with our theoretical analysis. An external magnetic field can be used to control the magnetic domains associated with the three ground-state wave vectors $\bm q_n$ along the cubic axes. Helices propagating along the field direction can easily acquire a net magnetization by transforming into a conical helix, while other helices would become anharmonic in field\,---\,this costs extra energy. As a result, the domain balance drastically shifts towards the conical domain. Cooling in field can select one type of domain completely and prepare the system in a single-domain state.

\begin{figure}
  \includegraphics[width=\columnwidth]{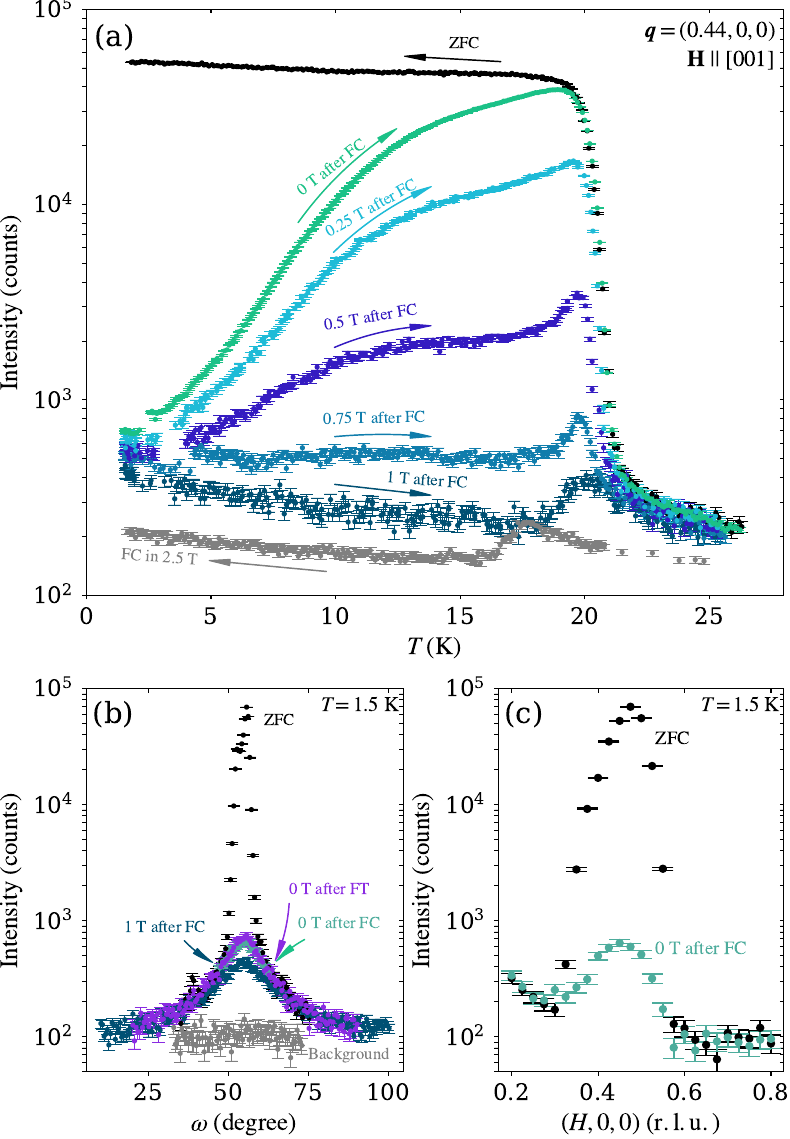}
	\caption{Elastic neutron scattering data for \zncr. Intensities are normalized to 30000 monitor counts. (a)~Temperature dependence of the suppressed magnetic reflection at $\bm q_{\text{PG}}$ in an external field. \enquote{FC} denotes \enquote{field cooling} in a 2.5~T field. (b)~Rocking scans at the position of the suppressed reflection at 1.5\,K after different protocols. The background reference curve is a rocking scan at 1.5\,K in 1\,T at $\bm q = (0.6, 0, 0)$. \enquote{FT} stands for \enquote{field training}\,---\,cooling in zero field followed by a field sweep up to 2.5\,T at low temperature. (c)~Scans along [100] direction with and without domain selection.}
	\label{fig:experiment}
\end{figure}

Although elastic neutron scattering measurements show that the applied field reliably drives the system into a single-domain configuration, there is still a significant intensity attributed to the suppressed ordering vector $\bm q_\text{PG}$, see \autoref{fig:experiment}. The residual intensity at $\bm q_\text{PG}$ is approximately two orders of magnitude weaker than the primary ordering peak at $\bm q_\text{B}$ but still several times stronger than the background [\autoref{fig:experiment}(b)]. The temperature behavior of the peak at $\bm q_\text{PG}$ was examined through a series of measurements on warming performed under different fields [\autoref{fig:experiment}(a)]. The zero field cooled (ZFC) curve serves as the main reference\,---\,it follows a normal order parameter behavior with a single transition at $T_\mathrm{N} = 21\,\mathrm{K}$. At non-zero fields, the intensity of the reflection is reduced relative to the ZFC curve due to field-induced domain selection. The domain balance at each point is determined by field and temperature conditions which control the equilibrium between propagation vectors and the relaxation rate (with thermal fluctuations favoring equal domain populations). At higher temperature, in the vicinity of the transition into the paramagnetic state, the intensity of the suppressed ordering vector exhibits a spike indicating that the thermal fluctuations prevail over field-induced domain selection causing the system to quickly relax back to the multi-domain state.

At lower temperatures, and especially in the presence of a magnetic field, the relaxation rate becomes extremely slow\,\cite{Inosov_2020}, and the temperature dependence is governed primarily by the evolution of the order parameter. It is clearest in the 1\,T field-warming and the 2.5\,T field-cooling curves where the intensity decreases smoothly with increasing temperature, following a slope similar to that of the ZFC curve (with minor deviations expected due to differing field conditions). It is apparent that when the field is lower than $\sim\!0.75$\,T the observed intensity reflects a mixture of relaxing domain selection and the remnant modulation at $\bm q_\text{PG}$, while at higher fields the relaxation does not take place, revealing the genuine temperature dependence of the order parameter at $\bm q_\text{PG}$.

The remnant peak is indeed magnetic: it varies with applied field and shows different intensities at 1\,T and in zero field. On the other hand, its intensity is independent of the sample's history and is identical for both field-cooling and field-training protocols [see \autoref{fig:experiment}(b)], indicating that the peak originates from an intrinsic equilibrium property of \zncr. A strong argument that this remnant peak is a distinct type of spin modulation in \zncr\ is the fact that its shape is drastically different from the primary reflection. The latter arises from long-range helical order and therefore appears resolution-limited\,---\,its width is determined primarily by instrumental resolution and sample mosaicity (FWHM is $\sim\!4^\circ$ in rocking-scan angle). In contrast, the remnant peak is much broader, exhibits a clear intrinsic width (FWHM is greater than $\sim\!16^\circ$) and displays a profile with Lorentzian-like tails, strongly suggesting its origin in short-range correlations. Notably, cuts along the ordering vector [\autoref{fig:experiment}(c)] reveal that both the long-range order peak and the short-range correlations are similar in terms of width along $\bm q_{\rm B}$, confirming that both modulations possess the same fixed periodicity. However, the short-range remnant peak has significant azimuthal broadening.

These observations indicate that \zncr\ realizes a genuine multi-$\bm q$ magnetic order, which suggests that the quartic modes are destabilized by negative harmonic mass terms. This is intriguing, since our theoretical analysis excludes this possibility for symmetry-respecting, bilinear interactions. Below 21~K, the crystal structure of \zncr\ is in fact known to lower its symmetry from cubic to tetragonal~\cite{zajdel_structure_2017}. This distortion is, however, tiny: the lattice parameters deviate from the cubic structure only by $\sigma=(\sqrt{2}a_{\rm tet} - c_\mathrm{tet})/(\sqrt{2}a_\mathrm{tet}) \sim\!8 \times 10^{-4}$. 
In general, this slightly lifts the degeneracy among spirals with wave vectors $\bm q_n$, either lowering or raising the energy of the $\bm q_\text{B}$ order relative to $\bm q_{\text{PG}n}$.
Respectively, these scenarios appear in our formalism as positive and negative mass terms. The second scenario, however, is inconsistent with an equilibrium magnetoelastic distortion~\cite{tchernyshyovSpinPeierlsPhasesPyrochlore2002}. The first scenario merely generates a small positive mass, which we estimate is of order $M \approx 5 \times 10^{-6}\alpha|J_1| N_u$~\cite{SM}, where $\alpha>0$ is an order-1 parameter for the dependence of the exchange couplings on the bond distance.
We conclude that the tetragonal distortion \textit{opposes}, rather than drives, the multi-$\bm{q}$ order.

Fluctuation effects are also an unlikely origin of the multi-$\bm q$ state, since order by disorder, of either thermal or quantum origin, typically selects coplanar states~\cite{Chalker1992,Reimers1993,Zhitomirsky2002,Hopkinson2007,Chernyshev2015}. The most natural explanation therefore lies in interactions beyond bilinear order, among which an isotropic biquadratic interaction is the most obvious candidate.

\textit{Role of biquadratic spin interactions.}---
To explain the observed multi-$\bm q$ state in \zncr, we add to $H$ a biquadratic interaction $\kappa (\bm S_{i\alpha}\cdot\bm S_{j\beta})^2$ on each nearest-neighbor bond $(i\alpha,j\beta)$ of the pyrochlore lattice, where $\kappa$ denotes the biquadratic coupling strength. Since this interaction is spin isotropic, the Goldstone modes remain unaffected. Repeating the calculation of the mass matrix $M$ in the space of quartic modes, we obtain~\cite{SM}
\begin{equation}
M=-32 N_u \kappa \sin^4\left(\frac{\xi a}{8}\right)\,\text{diag}(1,1,1,1),
\end{equation}
where $a$ is the cubic lattice constant. Hence, a {\it positive} biquadratic coupling $\kappa$ renders all quartic modes unstable and drives the system towards a multi-$\bm q$ state.

Neither our analytical considerations, nor our experiment, can
distinguish a thermal superposition of two double-$\bm{q}$ domains from a genuine triple-$\bm{q}$ order. We therefore perform classical Monte Carlo (simulated annealing) simulations on a frustrated fourth-neighbor $J_1,J_2,J_3,J_4$ Heisenberg model~\cite{Tymoshenko_2017}, supplemented by a biquadratic interaction $\kappa$.

\begin{figure}
    \centering
    \includegraphics[width=\linewidth]{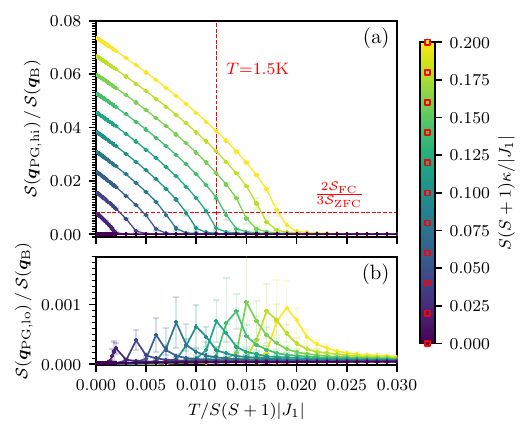}
    \caption{
    Monte Carlo elastic scattering
    strength at the pseudo-Goldstone momenta $\bm{q}_{\rm PG, hi}$ and $\bm{q}_{\rm
    PG, lo}$, plotted relative to the primary Bragg peak intensity $\cmcal{S}(\bm{q}_B)$ in
    the fourth-neighbor Heisenberg + biquadratic model. 
    Before averaging, each random seed is binned by which of $\bm{q}_{\rm PG, hi/lo}$ has greater intensity. Units are in terms of spin length $S=3/2$. We use the coupling constants $J_2/|J_1|=0.0118,~[J_3+\kappa S(S+1)/8]/|J_1|=0.164943,~J_4/|J_1|=-0.014$ corresponding to $\xi=4/9\; \text{r.l.u.}$; commensurate with the $L=18$ system size. Red, dashed crosshairs in panel (a) show the lowest experimental temperature, 1.5~K, and the approximate Bragg peak ratio seen in \autoref{fig:experiment}, corrected by a factor of 2/3 to account for domain coexistence.
   }
    \label{fig:numerics}
\end{figure}

In simulations, finite system size snaps $\bm{q}$ to a very coarse grid of allowed values, rendering a fine-grained scan of $\kappa$ alone impossible. It can be shown~\cite{SM} that within the single-$\bm{q}$ spiral phase, surfaces of constant 
 $\kappa S(S+1) + 8J_3$ have constant spiral wavevectors. Working in units of $J_1$, we held $J_2$ and $J_4$ fixed at the values previously determined from neutron scattering~\cite{Tymoshenko_2017} and varied $J_3$ along the fixed-$\bm{q}$ contour, ensuring a commensurate $\xi = 4/9$~r.l.u. order across the phase diagram.

 From \autoref{fig:numerics}, it can be seen that there is a phase transition from the high-temperature single-$\bm q$ state to a spontaneously chosen double-$\bm q$ state. The observed temperature dependence is consistent with a typical $\mathbb{Z}_2$ symmetry breaking, corresponding to the choice of pseudo-Goldstone mode. The small, noisy peaks of $\cmcal{S}(\bm q_{\rm PG, lo})$ near each $\kappa$'s critical temperature reveal critically enhanced diffuse spiral-mode fluctuations, which die down on both sides of the transition. 
It is not possible to clearly resolve this curve in \autoref{fig:experiment}(a) due to contamination by competing Bragg domains at higher $T$.

As an order-of-magnitude cross check, we plot the experimental temperature divided by $|J_1|=2.876$~meV~\cite{Tymoshenko_2017} and the Bragg peak ratio as crosshairs on \autoref{fig:numerics}(a). From this, we estimate $\kappa \simeq 0.1$~meV~\cite{SM}, within the range plausibly generated by exchange~\cite{Hoffmann2020}.

\textit{Discussion.}---
We have investigated under which conditions the quartic modes of planar spin spirals become unstable and drive the system towards multi-$\bm q$ order. On the pyrochlore lattice, no bilinear anisotropic interaction that respects the lattice symmetries can generate the required negative mass terms for a helimagnet. The multi-$\bm q$ order observed in \zncr\ by neutron scattering therefore calls for interactions beyond bilinear order, and we have identified a small isotropic biquadratic coupling as the most natural candidate. An instability of helimagnetic order of this type thus serves as a sensitive indicator of interactions beyond the bilinear level. We are not aware of other situations in which such interactions manifest so directly in the magnetic ground state.

The microscopic origin of the biquadratic interaction deserves further investigation, in particular its sign. Spin-lattice coupling, which is known to be sizable in chromium spinels~\cite{Rudolf2007,Gao2018}, typically generates a {\it negative} biquadratic coupling that favors collinear spin configurations~\cite{Penc2004,Bergman2006}. Within our analysis, such a term would stabilize the single-$\bm q$ spiral. The instability instead requires $\kappa>0$, which points to a different origin, such as higher-order exchange processes beyond the leading superexchange~\cite{Hoffmann2020}, possibly competing with the magnetoelastic contribution. Clarifying this mechanism by first-principles approaches would be an interesting task for future work.

Finally, our Monte Carlo simulations show that only one of the two quartic modes is spontaneously selected and acquires a finite amplitude, while the other remains soft. This has two important consequences. First, in a macroscopic sample one expects domains in which either $\bm q_{\text{PG}1}$ or $\bm q_{\text{PG}2}$ is selected, and their finite size naturally explains the observed enhanced width of the peak at $\bm q_\text{PG}$. Second, the double-$\bm q$ state remains compatible with the pseudo-Goldstone mode observed in our companion paper~\cite{Andriushin2026}. This mode requires an intact quartic mode, which persists at the unselected pseudo-Goldstone wave vector.

\bibliography{bibliography}

\smallskip
\noindent\textbf{Data availability.}
The data that support the findings of this article are publicly available\,\cite{4-01-1866}.

\noindent\textbf{Competing interests.}
The authors declare no competing interests.

\noindent\textbf{Acknowledgments.}
We acknowledge financial support from the German Research Foundation (DFG) through: individual grants PE~3318/3 and IN~209/9 (Project Nos.\ 455319354 and 434257385); through project C03 of the Collaborative Research Center SFB 1143 (project No.\ 247310070); through the W\"urzburg-Dresden Cluster of Excellence on Complexity and Topology in Quantum Matter\,---\,\textit{ct.qmat} (EXC~2147, Project No.~390858490); and through TRR\,360 (Project No.\ 492547816). V.\,T., L.{\kern.5pt}P., and V.\,F.\ acknowledge support through the National Agency for Research and Development (ANCD) of Moldova (project No.\ 011201). This work is based in part on experiments performed at the Institut Laue-Langevin (ILL), Grenoble, France. \smallskip

\onecolumngrid
\begin{center}
  \large\textbf{End Matter}
\end{center}
\twocolumngrid

\noindent\textit{Elastic neutron scattering.}\,---\,A mosaic of several ZnCr$_2$Se$_4$ single crystals with a total mass of $\sim\!0.4$~g was aligned with the cubic [001] direction vertical, giving access to the $(HK0)$ scattering plane. The neutron-scattering experiment was performed on the ThALES triple-axis spectrometer at the ILL, Grenoble, France~\cite{ThALES_Boehm2008}. We used a fixed final neutron wave number $k_\mathrm{f} = 1.150~\text{\AA}^{-1}$, a post-sample collimator and a curved analyzer. A 2.5-T cryomagnet was used to prepare a single-domain state and to measure the temperature dependence of the elastic neutron-scattering intensity in an applied field.

\clearpage
\onecolumngrid

\setcounter{equation}{0}
\setcounter{figure}{0}
\setcounter{table}{0}
\setcounter{section}{0}
\renewcommand{\theequation}{S\arabic{equation}}
\makeatletter\renewcommand{\fnum@figure}[1]{\figurename~S\thefigure.~}\makeatother
\makeatletter\renewcommand{\fnum@table}[1]{\tablename~S\thetable.~}\makeatother
\renewcommand{\thesection}{S\arabic{section}}
\renewcommand{\citenumfont}[1]{S#1}    %
\renewcommand{\bibnumfmt}[1]{[S#1]}    %

\begin{center}
  {\large\textbf{Supplemental Material for\\[2pt] ``Instability of single-${\it q}$ helimagnets as an indicator of non-Heisenberg interactions''}}\\[8pt]
\end{center}
\vspace{12pt}

\section{Luttinger-Tisza solvability}
We consider a spin-isotropic spiral spin system on a general lattice with $N_\text{sub}$ sites per unit cell labeled by a sublattice index $\alpha\in\{1,2,\ldots,N_\text{sub}\}$. Furthermore, $i$ labels the unit cell and ${\bm r}_{i\alpha}$ is the position of the site on sublattice $\alpha$ in the $i$th unit cell. The system has $N_u$ unit cells in total. The system's Heisenberg Hamiltonian is given by
\begin{equation}
H=\frac{1}{2}\sum_{ij}\sum_{\alpha\beta} J_{ij\alpha\beta}{\bm S}_{i\alpha}\cdot {\bm S}_{j\beta}.\label{eq:ham_SM}
\end{equation}
The factor of $1/2$ ensures that there is no double counting of terms when we sum over all indices independently. We treat the spins as classical variables with unit norm $\|{\bm S}_{i\alpha}\|=1$.

It will be convenient to work in Fourier space. Using the Fourier-transformed spins,
\begin{equation}
S_{i\alpha}^\mu=\frac{1}{\sqrt{N_u}}\sum_{\bm k} e^{\imath {\bm k}\cdot{\bm r}_{i\alpha}}S_\alpha^\mu(\bm k),\label{eq:ft}
\end{equation}
the Hamiltonian becomes 
\begin{equation}
H=\frac{1}{2}\sum_{\bm k}\sum_{\alpha\beta}\sum_\mu S_\alpha^\mu(\bm k)J_{\alpha\beta}({\bm k}) S_\beta^\mu(-{\bm k}),\label{eq:hamk}
\end{equation}
where
\begin{equation}
J_{\alpha\beta}({\bm k})=\sum_i e^{\imath {\bm k}\cdot ({\bm r}_{i\alpha}-{\bm r}_{j\beta})}J_{ij\alpha\beta},\label{eq:ham_matrix}
\end{equation}
and $\mu\in\{x,y,z\}$ are the spin components. In Eq.~(\ref{eq:ham_matrix}), $j$ denotes an arbitrary fixed reference unit cell. We assume that the lattice is inversion symmetric for every site such that $J_{\alpha\beta}({\bm k})$ is a real symmetric matrix with $J_{\alpha\beta}({\bm k})=J_{\alpha\beta}(-{\bm k})$.

We apply the Luttinger-Tisza method, which constructs ground states by diagonalizing $J_{\alpha\beta}({\bm k})$ and finding the lowest eigenvalue $\epsilon_0$ among all wave vectors $\bm k$. We denote the corresponding eigenvector by $s_\alpha^0$ and the ground state wave vectors by $\bm q_n$, i.e.,
\begin{equation}\label{eq:eigenequation}
\sum_\beta J_{\alpha\beta}({\bm q}_n)s^0_\beta=\epsilon_0 s_\alpha^0,
\end{equation}
The condition for $s_\alpha^0$ to represent the exact ground state is that all its components have the same norm, $|s^0_\alpha|=1$. This condition is equivalent to the requirement that the spins on all lattice sites are normalized to the same value. Since the eigenvectors can always be chosen real, the solvability condition can be restated as $s^0_\alpha=\pm1$. By a sublattice-dependent transformation of the spin order that flips the spin axes on the sublattices with $s^0_\alpha=-1$, one may assume $s^0_\alpha=+1$ for all components without loss of generality.

In this case, we obtain an exact ground state as
\begin{equation}
\bm S^0_{\alpha}(\bm k)=\frac{\sqrt{N_u}}{2}\left(\begin{array}{c}
\delta_{\bm k,\bm q_n}+\delta_{\bm k,-\bm q_n}\\
\frac{1}{\imath}(\delta_{\bm k,\bm q_n}-\delta_{\bm k,-\bm q_n})\\
0
\end{array}\right), \label{eq:gsk}
\end{equation}
where, without loss of generality, we have chosen the $x$-$y$ plane as the spiral plane. Note the absence of any $\alpha$-dependence in Eq.~(\ref{eq:gsk}) due to $s^0_\alpha=+1$. The superposition of spin configurations with $\bm q_n$ and $-{\bm q}_n$ in Eq.~(\ref{eq:gsk}) is needed to obtain a real-valued spin state. Inserting Eq.~(\ref{eq:gsk}) into Eq.~(\ref{eq:ft}) and allowing for an arbitrary in-plane phase $\varphi$, the ground states in real space have the form of single-$\bm q$ planar spirals
\begin{equation}
{\bm S}^0_{i\alpha}=\left(\begin{array}{c}
\cos({\bm q}_n\cdot{\bm r}_{i\alpha}+\varphi)\\
\sin({\bm q}_n\cdot{\bm r}_{i\alpha}+\varphi)\\
0
\end{array}\right).\label{eq:gs}
\end{equation}
The ground state energy is given by
\begin{equation}
E_0=\frac{N_u}{2}\sum_{\alpha\beta} J_{\alpha\beta}(\bm q_n)=\frac{N_u N_\text{sub}}{2}\epsilon_0.\label{eq:e0}
\end{equation}

There is usually more than one pair of inequivalent ground state wave vectors $\pm\bm q_n$ which are related by lattice symmetries. An important property which we will exploit below is that Eq.~(\ref{eq:gs}) is a ground state at all symmetry-related ground state wave vectors $\bm q_{n}$ (which implies that $s_\alpha^0=+1$ is a Luttinger-Tisza eigenvector at all ground state wave vectors $\bm q_n$). If no sublattice-dependent spin flip transformation is required to eliminate eigenvector components $s_\alpha^0=-1$, this is always guaranteed because lattice symmetry transformations only permute the sublattice indices in Eq.~(\ref{eq:gs}) but otherwise do not change the spin structure. A subtlety, however, can arise when sublattice transformations are applied to yield $s_\alpha^0=+1$ at a given ground state wave vector $\bm q_n$, because this might not simultaneously eliminate components $s_\alpha^0=-1$ at all other symmetry related ground state wave vectors. These systems must be excluded in the following considerations. 

All these conditions are fulfilled for the pyrochlore model ($N_\text{sub}=4$) for ZnCr$_2$Se$_4$ proposed in Ref.~\citeS{PG2017PRX} which contains pyrochlore couplings between spins up to fourth nearest neighbors, yielding single-$\bm q$ ground state spirals as in Eq.~(\ref{eq:gs}). The spiral ground state wave vectors $\bm q_n$ for this model are along the cubic axes, $\bm q_n=(\pm\xi,0,0)$, $(0,\pm\xi,0)$, $(0,0,\pm\xi)$.

\section{Identification of soft deformations}

We now slightly perturb the single-$\bm q$ state in Eq.~(\ref{eq:gs}) and calculate the energy of these deformations. We consider the perturbation in the local frame of spin directions, writing the perturbed spin state as
\begin{equation}
{\bm S}_{i\alpha}=\left(\begin{array}{ccc}
\cos({\bm q}_n\cdot{\bm r}_{i\alpha}+\varphi)&-\sin({\bm q}_n\cdot{\bm r}_{i\alpha}+\varphi)&0\\
\sin({\bm q}_n\cdot{\bm r}_{i\alpha}+\varphi)&\cos({\bm q}_n\cdot{\bm r}_{i\alpha}+\varphi)&0\\
0&0&1
\end{array}\right)
\left[\left(\begin{array}{c}1\\0\\0
\end{array}\right)+\left(\begin{array}{c}0\\\Delta^\parallel_{i\alpha}\\\Delta^\perp_{i\alpha}
\end{array}\right)\right]\frac{1}{\sqrt{1+(\Delta^\parallel_{i\alpha})^2+(\Delta^\perp_{i\alpha})^2}}.\label{eq:perturbation}
\end{equation}
Here, the second vector in the square bracket is the perturbation, where $\Delta^\parallel_{i\alpha}$ ($\Delta^\perp_{i\alpha}$) describes a deformation in the spiral plane (perpendicular to the spiral plane) and the last factor in Eq.~(\ref{eq:perturbation}) is the normalization. The rotation matrix in Eq.~(\ref{eq:perturbation}) transforms the spins into the global coordinate frame.

Expanding the normalization in Eq.~(\ref{eq:perturbation}) up to quadratic order yields
\begin{equation}
{\bm S}_{i\alpha}=\left(\begin{array}{ccc}
\cos({\bm q}_n\cdot{\bm r}_{i\alpha}+\varphi)&-\sin({\bm q}_n\cdot{\bm r}_{i\alpha}+\varphi)&0\\
\sin({\bm q}_n\cdot{\bm r}_{i\alpha}+\varphi)&\cos({\bm q}_n\cdot{\bm r}_{i\alpha}+\varphi)&0\\
0&0&1
\end{array}\right)
\left[\left(\begin{array}{c}1\\0\\0
\end{array}\right)+\left(\begin{array}{c}-\frac{1}{2}[(\Delta^\parallel_{i\alpha})^2+(\Delta^\perp_{i\alpha})^2]\\\Delta^\parallel_{i\alpha}\\\Delta^\perp_{i\alpha}
\end{array}\right)\right]+\mathcal{O}((\Delta^{\parallel/\perp})^3).\label{eq:perturbation2}
\end{equation}
Inserting Eq.~(\ref{eq:perturbation2}) into Eq.~(\ref{eq:ham_SM}) and using the fact that the terms linear in the perturbation must vanish, one obtains to quadratic order
\begin{align}
H= E_0+\frac{1}{2}\sum_{ij}\sum_{\alpha\beta}J_{ij\alpha\beta}&\left[-\frac{1}{2}\left((\Delta^\parallel_{i\alpha})^2+(\Delta^\perp_{i\alpha})^2+(\Delta^\parallel_{j\beta})^2+(\Delta^\perp_{j\beta})^2\right)\cos\left[{\bm q}_n\cdot({\bm r}_{i\alpha}-{\bm r}_{j\beta})\right]\right.\notag\\
&\left.+\Delta^\parallel_{i\alpha}\Delta^\parallel_{j\beta}\cos\left[{\bm q}_n\cdot({\bm r}_{i\alpha}-{\bm r}_{j\beta})\right]+\Delta^\perp_{i\alpha}\Delta^\perp_{j\beta}\right]+\mathcal{O}((\Delta^{\parallel/\perp})^3).
\end{align}
A simplification can be made in the first line, yielding
\begin{align}
H&= E_0+\frac{1}{2}\sum_{ij}\sum_{\alpha\beta}J_{ij\alpha\beta}\left[-\left((\Delta^\parallel_{i\alpha})^2+(\Delta^\perp_{i\alpha})^2\right)\cos\left[{\bm q}_n\cdot({\bm r}_{i\alpha}-{\bm r}_{j\beta})\right]+\Delta^\parallel_{i\alpha}\Delta^\parallel_{j\beta}\cos\left[{\bm q}_n\cdot({\bm r}_{i\alpha}-{\bm r}_{j\beta})\right]+\Delta^\perp_{i\alpha}\Delta^\perp_{j\beta}\right]\notag\\
&+\mathcal{O}((\Delta^{\parallel/\perp})^3)\notag\\
&=E_0+\frac{1}{2}\sum_{ij}\sum_{\alpha\beta}\left(J_{ij\alpha\beta}^\parallel(\bm q_n)\Delta_{i\alpha}^\parallel \Delta_{j\beta}^\parallel+J_{ij\alpha\beta}^\perp(\bm q_n)\Delta_{i\alpha}^\perp \Delta_{j\beta}^\perp\right)+\mathcal{O}((\Delta^{\parallel/\perp})^3),
\end{align}
where we have defined
\begin{align}
J_{ij\alpha\beta}^\parallel(\bm k)&=-\delta_{ij}\delta_{\alpha\beta}\sum_{l}\sum_{\gamma}J_{il\alpha\gamma}\cos\left[{\bm k}\cdot({\bm r}_{i\alpha}-{\bm r}_{l\gamma})\right]+J_{ij\alpha\beta}\cos\left[{\bm k}\cdot({\bm r}_{i\alpha}-{\bm r}_{j\beta})\right],\notag\\
J_{ij\alpha\beta}^\perp(\bm k)&=-\delta_{ij}\delta_{\alpha\beta}\sum_{l}\sum_{\gamma}J_{il\alpha\gamma}\cos\left[{\bm k}\cdot({\bm r}_{i\alpha}-{\bm r}_{l\gamma})\right]+J_{ij\alpha\beta}.\label{eq:jnew}
\end{align}
The remaining task is to diagonalize $J_{ij\alpha\beta}^\parallel(\bm k)$ and $J_{ij\alpha\beta}^\perp(\bm k)$, which are the Hessians for in-plane and out-of-plane fluctuations. By construction, the Hessians are positive semidefinite and we are particularly interested in the zero modes at this harmonic order. Note that the spin normalization has already been taken care of up to quadratic order in Eq.~(\ref{eq:perturbation2}). Therefore, in contrast to the diagonalization of $J_{\alpha\beta}(\bm k)$, there is no condition on the normalization of individual entries of the eigenvectors of $J_{ij\alpha\beta}^\parallel(\bm k)$ and $J_{ij\alpha\beta}^\perp(\bm k)$. This means that our approach of finding the eigenmodes of spin deformations of spiral ground states is {\it exact} up to quadratic order.

In analogy to Eq.~(\ref{eq:hamk}), we block-diagonalize the Hessians via a Fourier transformation,
\begin{equation}
\Delta_{i\alpha}^{\parallel/\perp}=\frac{1}{\sqrt{N_u}}\sum_{\bm k} e^{\imath {\bm k}\cdot{\bm r}_{i\alpha}}\Delta_\alpha^{\parallel/\perp}(\bm k),\label{eq:ftdelta}
\end{equation}
such that
\begin{equation}
H=E_0+\frac{1}{2}\sum_{\bm k}\sum_{\alpha\beta}\left(\Delta^{\parallel}_\alpha(\bm k)J^\parallel_{\alpha\beta}({\bm k},{\bm q}_n)\Delta^{\parallel}_\beta(-{\bm k})+\Delta^{\perp}_\alpha(\bm k)J^\perp_{\alpha\beta}({\bm k},{\bm q}_n)\Delta^{\perp}_\beta(-{\bm k})\right)+\mathcal{O}((\Delta^{\parallel/\perp})^3)
\end{equation}
with
\begin{equation}
J^{\parallel/\perp}_{\alpha\beta}({\bm k},{\bm q})=\sum_i e^{\imath {\bm k}\cdot ({\bm r}_{i\alpha}-{\bm r}_{j\beta})}J^{\parallel/\perp}_{ij\alpha\beta}(\bm q).\label{eq:ham_matrix2}
\end{equation}
Note the distinction between the two wave vector arguments $\bm k$ and $\bm q$ in $J^{\parallel/\perp}_{\alpha\beta}({\bm k},{\bm q})$, where $\bm q$ is the wave vector of the unperturbed planar spiral and $\bm k$ is the wave vector of the deformation of this spiral.

Inserting Eqs.~(\ref{eq:jnew}) into Eq.~(\ref{eq:ham_matrix2}) and using Eq.~(\ref{eq:ham_matrix}) one finds
\begin{align}
J_{\alpha\beta}^\parallel({\bm k},{\bm q}_n)&=-\frac{1}{2}\delta_{\alpha\beta}\sum_\gamma(J_{\alpha\gamma}(\bm q_n)+J_{\alpha\gamma}(-{\bm q}_n))+\frac{1}{2}(J_{\alpha\beta}({\bm k}+{\bm q}_n)+J_{\alpha\beta}({\bm k}-{\bm q}_n)),\notag\\
J_{\alpha\beta}^\perp({\bm k},{\bm q}_n)&=-\frac{1}{2}\delta_{\alpha\beta}\sum_\gamma(J_{\alpha\gamma}(\bm q_n)+J_{\alpha\gamma}(-{\bm q}_n))+J_{\alpha\beta}({\bm k}).
\end{align}
Using Eq.~(\ref{eq:eigenequation}) with $s_\alpha^0=1$ and $J_{\alpha\beta}(\bm k)=J_{\alpha\beta}(-\bm k)$ we finally arrive at
\begin{align}
J_{\alpha\beta}^\parallel({\bm k},{\bm q}_n)&=-\delta_{\alpha\beta}\epsilon_0+\frac{1}{2}(J_{\alpha\beta}({\bm k}+{\bm q}_n)+J_{\alpha\beta}({\bm k}-{\bm q}_n)),\label{eq:hessian_k1}\\
J_{\alpha\beta}^\perp({\bm k},{\bm q}_n)&=-\delta_{\alpha\beta}\epsilon_0+J_{\alpha\beta}({\bm k}).\label{eq:hessian_k2}
\end{align}
Diagonalizing $J_{\alpha\beta}^\parallel({\bm k},{\bm q}_n)$ and $J_{\alpha\beta}^\perp({\bm k},{\bm q}_n)$ and finding the smallest eigenvalues is straightforward because we already know that $J_{\alpha\beta}(\bm k)$ on the right-hand sides of Eqs.~(\ref{eq:hessian_k1}) and (\ref{eq:hessian_k2}) acquires the smallest eigenvalue $\epsilon_0$ when $\bm k$ is one of the ground state wave vectors $\bm q_n$.

We can directly read off the zero modes from Eqs.~(\ref{eq:hessian_k1}) and (\ref{eq:hessian_k2}). Starting with the in-plane fluctuations in Eq.~(\ref{eq:hessian_k1}), the only zero mode is obtained for ${\bm k}=0$. This is a Goldstone mode which corresponds to a global in-plane rotation of the spiral.

The out-of-plane fluctuations in Eq.~(\ref{eq:hessian_k2}) have zero modes whenever $\bm k$ is one of the ground state wave vectors $\bm q_n$. Denoting the ground state wave vector $\bm q_n$ selected in the unperturbed planar spiral $\bm q_\text{B}$ (``B'' stands for Bragg), the wave vector of the deformation $\bm k$ can either be the same ground state wave vector $\bm k=\pm\bm q_\text{B}$ (with an irrelevant sign) or a different one $\bm k\neq\pm\bm q_\text{B}$. We call these other ground state wave vectors $\bm q_{\text{PG}n}$, where ``PG'' stands for pseudo-Goldstone.

The case $\bm k=\pm\bm q_\text{B}$ leads to two more Goldstone modes which correspond to rotations of the spiral plane about an axis in the $x$-$y$ plane. Together with the in-plane Goldstone mode at $\bm k=0$ from Eq.~(\ref{eq:hessian_k1}) this yields three Goldstone modes in total, which is the expected number for a type of magnetic order that breaks all continuous spin rotation symmetries. By superposition of modes with ${\bm k}=+{\bm q}_\text{B}$ and ${\bm k}=-{\bm q}_\text{B}$ we can construct two {\it real} out-of-plane Goldstone modes which, according to Eq.~(\ref{eq:ftdelta}) read
\begin{equation}
\Delta_{i\alpha}^{\perp}\propto\sum_{\bm k} e^{\imath {\bm k}\cdot{\bm r}_{i\alpha}}(\delta_{{\bm k},{\bm q}_\text{B}}+\delta_{{\bm k},-{\bm q}_\text{B}})\propto \cos({\bm q}_\text{B}\cdot {\bm r}_{i\alpha}),\label{eq:goldstone_modes1}
\end{equation}
and
\begin{equation}
\Delta_{i\alpha}^{\perp}\propto\sum_{\bm k} e^{\imath {\bm k}\cdot{\bm r}_{i\alpha}}(\delta_{{\bm k},{\bm q}_\text{B}}-\delta_{{\bm k},-{\bm q}_\text{B}})\propto \sin({\bm q}_\text{B}\cdot {\bm r}_{i\alpha}).\label{eq:goldstone_modes2}
\end{equation}

Importantly, there are additional out-of-plane zero modes from $J_{\alpha\beta}^\perp({\bm k},{\bm q}_\text{B})$ when ${\bm k}=\pm{\bm q}_{\text{PG}n}$. Again, and in analogy to Eqs.~(\ref{eq:goldstone_modes1}) and (\ref{eq:goldstone_modes2}), for each $\bm q_{\text{PG}n}$, by superposition of modes with ${\bm k}={\bm q}_{\text{PG}n}$ and ${\bm k}=-{\bm q}_{\text{PG}n}$ we can construct two {\it real} zero modes given by
\begin{equation}
\Delta_{i\alpha}^{\perp}\propto \cos({\bm q}_{\text{PG}n}\cdot {\bm r}_{i\alpha})\quad\text{and}\quad\Delta_{i\alpha}^{\perp}\propto \sin({\bm q}_{\text{PG}n}\cdot {\bm r}_{i\alpha}).\label{eq:quartic_mode}
\end{equation}
Note that, apart from the $\alpha$-dependence in $\bm r_{i\alpha}$, there is no additional sublattice-dependence in Eq.~(\ref{eq:quartic_mode}) such as a sublattice-dependent modulation of the amplitudes. This is because the Luttinger-Tisza eigenvector $s_\alpha^0=+1$ of $J_{\alpha\beta}(\bm k)$ is a ground state eigenvector at all momenta, $\pm\bm q_\text{B}$ and $\pm\bm q_{\text{PG}n}$.

The two zero modes in Eq.~(\ref{eq:quartic_mode}) exist for all pairs of pseudo-Goldstone wave vectors $\pm\bm q_{\text{PG}n}$. In the pyrochlore lattice case where the ground state wave vectors are along the cubic axes, this gives rise to two pairs $\pm\bm q_\text{PG1}$ and $\pm\bm q_\text{PG2}$, thus, two pairs of modes as in Eq.~(\ref{eq:quartic_mode}). These deformations are not Goldstone modes; therefore, higher orders in the perturbation series usually do not vanish. In the generic case these are {\it quartic} modes. (For simplicity, we call them {\it quartic} in the following even though they might be of higher order.) Note that the existence of these quartic modes is only guaranteed if the ground states are of the single-$\bm q$ form as in Eq.~(\ref{eq:gs}).

\section{Bilinear anisotropic perturbations}
Since the quartic modes in Eq.~(\ref{eq:quartic_mode}) are accidental, one expects that the addition of a generic non-Heisenberg interaction will introduce a non-zero harmonic term, either with a positive or a negative prefactor---the latter makes the unperturbed planar spiral unstable. Here we add a general bilinear anisotropic perturbation $V$ and investigate how it acts in the subspace of quartic modes. The full Hamiltonian is given by $H+V$ where $V$ has the general form
\begin{equation}
V=\frac{1}{2}\sum_{ij}\sum_{\alpha\beta}\sum_{\mu\nu}S^\mu_{i\alpha} V^{\mu\nu}_{ij\alpha\beta} S^\nu_{j\beta},\label{eq:ham_aniso_supp}
\end{equation}
and $V^{\mu\nu}_{ij\alpha\beta}$ are some anisotropic couplings.

Strictly speaking, the analysis in the space of quartic modes only captures deformations at the ground-state wave vectors $\bm q_n$ of $H$. Deformation modes of $H$ at nearby wave vectors $\bm k$ are not zero modes, but their mass vanishes as $\bm k\to\bm q_n$ and is therefore small. Since this mass is of the same order as the quadratic contribution of a weak perturbation $V$ at the same wave vectors, the two compete directly and may change the sign of the total mass in $H+V$. An instability could thus, in principle, occur at a wave vector close to, but different from, $\bm q_n$ which modifies the ground state wave vectors. If $V$ respects the lattice symmetries, the direction of $\bm q_n$ is typically fixed by symmetry and only its length changes, i.e., all ground-state wave vectors are rescaled by a common constant, $\bm q_n\rightarrow \zeta\bm q_n$. Provided that the Luttinger-Tisza eigenvector $s_\alpha^0=+1$ remains the lowest eigenvector of $J_{\alpha\beta}(\bm k)$ at the renormalized wave vectors, our analysis in the space of quartic modes then applies at these wave vectors.

The deformations corresponding to the quartic modes are parametrized as
\begin{equation}
\Delta_{i\alpha}^\perp=a_{\bm q_\text{PG1}} \cos({\bm q}_\text{PG1}\cdot{\bm r}_{i\alpha})+b_{\bm q_\text{PG1}} \sin({\bm q}_\text{PG1}\cdot{\bm r}_{i\alpha})+a_{\bm q_\text{PG2}} \cos({\bm q}_\text{PG2}\cdot{\bm r}_{i\alpha})+b_{\bm q_\text{PG2}} \sin({\bm q}_\text{PG2}\cdot{\bm r}_{i\alpha}),\label{eq:parametrize_soft}
\end{equation}
and we define $\bm \delta=(a_{\bm q_\text{PG1}},b_{\bm q_\text{PG1}},a_{\bm q_\text{PG2}},b_{\bm q_\text{PG2}})$ as the vector of the soft mode amplitudes. The goal is to calculate $V$ in the space of these amplitudes up to harmonic order, i.e., we expand
\begin{equation}\label{eq:expand}
V=V_0+\bm m\cdot \bm \delta+\frac{1}{2}\bm \delta^\text{T} M \bm\delta,
\end{equation}
with the 4-dimensional vector $\bm m$ and the $4\times 4$ matrix $M$. Here, $V_0$ is the energy of the planar spiral $\bm S^0_{i\alpha}$ resulting from the anisotropic perturbation. Furthermore, $\bm m$ is a drift term of the quartic-mode amplitudes, generating them directly in linear order. For translation-invariant $V$, however, $\bm m=0$: to linear order, the quartic modes only add spin components with wave vectors $\pm\bm q_{\text{PG}n}$, while the undeformed spiral only has spin components with wave vectors $\pm\bm q_\text{B}$. Since a translation-invariant $V$ only couples spin components with opposite wave vectors [see Eq.~(\ref{eq:ham_anisok}) below], a term linear in $\bm\delta$ cannot occur. Of most interest are the eigenvalues of $M$, which are the mass terms acquired by the pseudo-Goldstone modes due to the perturbation $V$.

We also note that in harmonic order, the Goldstone modes do not couple to the quartic modes. Specifically, a mixed term such as $\sim\!a_{\bm q_\text{B}}a_{\bm q_\text{PG1}}$ in an expansion of $V$ in the space of soft modes, where $a_{\bm q_\text{B}}$ is the amplitude of the Goldstone mode in Eq.~(\ref{eq:goldstone_modes1}), cannot appear. The reason is the same as for $\bm m=0$: the Goldstone modes only involve spin components with wave vectors $\pm\bm q_\text{B}$ (or $\bm k=0$ in the local frame, which corresponds to $\pm\bm q_\text{B}$ in the global frame), and these cannot be coupled to the components at $\pm\bm q_{\text{PG}n}$ by a translation-invariant $V$. Therefore, we may investigate the effect of the perturbation $V$ in the space of quartic modes independently of the effect of $V$ on the Goldstone modes.

To calculate $M$, we insert $\Delta_{i\alpha}^\perp$ from Eq.~(\ref{eq:parametrize_soft}) into Eq.~(\ref{eq:perturbation2}), resulting in spin configurations $\bm S_{i\alpha}^\mu$ which we then insert into Eq.~(\ref{eq:ham_aniso_supp}). Without loss of generality, we assume the spiral plane in Eq.~(\ref{eq:perturbation2}) to be the $x$-$y$ plane and we set $\varphi=0$. The calculation is best performed in Fourier space where
\begin{equation}
V=\frac{1}{2}\sum_{\bm k}\sum_{\alpha\beta}\sum_{\mu\nu} S_\alpha^\mu(\bm k)V^{\mu\nu}_{\alpha\beta}({\bm k}) S_\beta^\nu(-{\bm k}).\label{eq:ham_anisok}
\end{equation}
Here, $S^\mu_\alpha(\bm k)$ is defined as in Eq.~(\ref{eq:ft}) and
\begin{equation}
V^{\mu\nu}_{\alpha\beta}({\bm k})=\sum_i e^{\imath {\bm k}\cdot ({\bm r}_{i\alpha}-{\bm r}_{j\beta})}V^{\mu\nu}_{ij\alpha\beta}.\label{eq:v_matrix}
\end{equation}
Importantly, the considered spin configuration $\bm S_{i\alpha}$ depends on the sublattice index $\alpha$ only through products $\bm k\cdot\bm r_{i\alpha}$, as they appear in the Fourier transform in Eq.~(\ref{eq:ft}). This means that $S_{\alpha}^\mu(\bm k)$ has in fact no explicit dependence on $\alpha$ such that we can write $S_\alpha^\mu(\bm k)\equiv S^\mu(\bm k)$ which simplifies Eq.~(\ref{eq:ham_anisok}),
\begin{equation}
V=\frac{1}{2}\sum_{\bm k}\sum_{\mu\nu} S^\mu(\bm k) S^\nu(-{\bm k})\sum_{\alpha\beta}V^{\mu\nu}_{\alpha\beta}({\bm k})=\frac{1}{2}\sum_{\bm k}\sum_{\mu\nu} S^\mu(\bm k) V^{\mu\nu}(\bm k)S^\nu(-{\bm k}),\label{eq:ham_anisok2}
\end{equation}
where
\begin{equation}
V^{\mu\nu}(\bm k)=\sum_{\alpha\beta}V^{\mu\nu}_{\alpha\beta}({\bm k}).\label{eq:subl_sum}
\end{equation}

Since we consider inversion-symmetric lattices, $V^{\mu\nu}(\bm k)$ must be real, $V^{\mu\nu}(\bm k)=(V^{\mu\nu}(\bm k))^*$, and symmetric under wave vector inversion, $V^{\mu\nu}(\bm k)=V^{\mu\nu}(-\bm k)$. Furthermore, without making further symmetry assumptions, one finds
\begin{equation}
V^{\mu\nu}(\bm k)=\sum_i\sum_{\alpha\beta} e^{\imath {\bm k}\cdot ({\bm r}_{i\alpha}-{\bm r}_{j\beta})}V^{\mu\nu}_{ij\alpha\beta}=\sum_j\sum_{\alpha\beta} e^{-\imath {\bm k}\cdot ({\bm r}_{i\alpha}-{\bm r}_{j\beta})}V^{\mu\nu}_{ji\beta\alpha}=\sum_j\sum_{\alpha\beta} e^{-\imath {\bm k}\cdot ({\bm r}_{i\alpha}-{\bm r}_{j\beta})}V_{ij\alpha\beta}^{\nu\mu}=V^{\nu\mu}(-\bm k)=V^{\nu\mu}(\bm k).\label{eq:property2}
\end{equation}
In the second step we have made the replacements $i\leftrightarrow j$, $\alpha\leftrightarrow \beta$ and in the third step we have used $V_{ij\alpha\beta}^{\mu\nu}=V^{\nu\mu}_{ji\beta\alpha}$. This means that for each wave vector $\bm k$, six independent real components of $V^{\mu\nu}(\bm k)$ can contribute to $M$ which are given by
\begin{equation}
V^{xx}(\bm k),\;V^{yy}(\bm k),\;V^{zz}(\bm k),\;V^{xy}(\bm k),\;V^{xz}(\bm k),\;V^{yz}(\bm k).\label{eq:ninev}
\end{equation}
Again, due to translation invariance, $M$ must be diagonal.
To calculate $M$, it suffices to keep terms up to second order in $\bm\delta$. With $\theta_{i\alpha}=\bm q_\text{B}\cdot\bm r_{i\alpha}$ and $\Delta^\parallel_{i\alpha}=0$, Eq.~(\ref{eq:perturbation2}) gives
\begin{equation}
\bm S_{i\alpha}=\left(\begin{array}{c}
(1+c_{i\alpha})\cos\theta_{i\alpha}\\
(1+c_{i\alpha})\sin\theta_{i\alpha}\\
\Delta^\perp_{i\alpha}
\end{array}\right),\quad c_{i\alpha}=-\frac{1}{2}(\Delta^\perp_{i\alpha})^2,\label{eq:spin_soft}
\end{equation}
with $\Delta^\perp_{i\alpha}$ from Eq.~(\ref{eq:parametrize_soft}). Inserting this into Eq.~(\ref{eq:ham_anisok2}), there are three types of terms up to second order in $\bm\delta$:

(i) Terms coupling $S^z$ to $S^x$ or $S^y$ are linear in $\bm\delta$ to lowest order. As argued above for $\bm m$, they vanish because $S^z(\bm k)$ is non-zero only at $\bm k=\pm\bm q_\text{PG1},\pm\bm q_\text{PG2}$, while the in-plane components $S^{x/y}(\bm k)$ are non-zero only at $\bm k=\pm\bm q_\text{B}$ in zeroth order.

(ii) The term $\sim\!V^{zz}$. Writing $a_n\equiv a_{\bm q_{\text{PG}n}}$ and $b_n\equiv b_{\bm q_{\text{PG}n}}$ ($n=1,2$), the Fourier transform of $S^z_{i\alpha}=\Delta^\perp_{i\alpha}$ reads
\begin{equation}
S^z(\bm k)=\frac{\sqrt{N_u}}{2}\sum_{n=1,2}\left[(a_n-\imath b_n)\delta_{\bm k,\bm q_{\text{PG}n}}+(a_n+\imath b_n)\delta_{\bm k,-\bm q_{\text{PG}n}}\right].
\end{equation}
Inserting this into Eq.~(\ref{eq:ham_anisok2}) and using $V^{zz}(\bm k)=V^{zz}(-\bm k)$ yields
\begin{equation}
V_{zz}=\frac{N_u}{4}\sum_{n=1,2}V^{zz}(\bm q_{\text{PG}n})\left(a_n^2+b_n^2\right).
\end{equation}

(iii) Terms involving the in-plane components. We call their sum $V_\parallel$. Denoting by $\bm S^0_{i\alpha}$ the undeformed spiral and by $\delta\bm S_{i\alpha}=c_{i\alpha}\bm S^0_{i\alpha}$ the second-order correction of the in-plane components in Eq.~(\ref{eq:spin_soft}), one finds in second order, using $V^{\mu\nu}(\bm k)=V^{\nu\mu}(-\bm k)$,
\begin{equation}
V_\parallel=\sum_{\bm k}\sum_{\mu\nu\in\{x,y\}}S^{0,\mu}(\bm k)V^{\mu\nu}(\bm k)\,\delta S^\nu(-\bm k).
\end{equation}
Since $S^{0,\mu}(\bm k)$ is non-zero only at $\bm k=\pm\bm q_\text{B}$, only the components of $\delta S^\nu(\bm k)$ at $\bm k=\mp\bm q_\text{B}$ contribute. Writing $(\Delta^\perp_{i\alpha})^2=\frac{1}{2}\sum_n(a_n^2+b_n^2)+\ldots$, where the dots denote terms oscillating with wave vectors $\pm2\bm q_{\text{PG}n}$ and $\pm\bm q_\text{PG1}\pm\bm q_\text{PG2}$, only the constant part of $c_{i\alpha}$ leaves the wave vectors of $\bm S^0_{i\alpha}$ unchanged and contributes. Hence, effectively $\delta\bm S_{i\alpha}=\bar c\,\bm S^0_{i\alpha}$ with $\bar c=-\frac{1}{4}\sum_n(a_n^2+b_n^2)$, and
\begin{equation}
V_\parallel=2\bar c\,V_0=-\frac{1}{2}V_0\sum_{n=1,2}\left(a_n^2+b_n^2\right),
\end{equation}
where $V_0=\frac{N_u}{4}\left(V^{xx}(\bm q_\text{B})+V^{yy}(\bm q_\text{B})\right)$ is the energy of the undeformed spiral [see Eq.~(\ref{eq:v0}) below].

Adding (ii) and (iii) gives $V=V_0+\frac{1}{2}\sum_n\left[\frac{N_u}{2}V^{zz}(\bm q_{\text{PG}n})-V_0\right](a_n^2+b_n^2)$. Comparing with Eq.~(\ref{eq:expand}), one finds
\begin{equation}
M=-\frac{N_u}{4}\left(V^{xx}(\bm q_\text{B})+V^{yy}(\bm q_\text{B})\right)\text{diag}(1,1,1,1)
+\frac{N_u}{2}\,\text{diag}\left(V^{zz}(\bm q_{\text{PG1}}),V^{zz}(\bm q_{\text{PG1}}),V^{zz}(\bm q_{\text{PG2}}),V^{zz}(\bm q_{\text{PG2}})\right).\label{eq:quartic_mass}
\end{equation}

\section{Application to helical and cycloidal spirals with cubic symmetries}
To investigate the implications of the mass terms of the quartic modes in Eq.~(\ref{eq:quartic_mass}) we consider the cases of helical and cycloidal spirals with cubic symmetries as realized on the pyrochlore lattice. Which of the two cases is realized is decided by $V_0$ in Eq.~(\ref{eq:expand}) which selects the energetically preferred spiral plane.

Specifically, for the calculation of $V_0$, we Fourier-transform the unperturbed $x$-$y$ spiral in Eq.~(\ref{eq:gs}), which yields
\begin{equation}
S_\alpha^\mu(\bm k)=\frac{\sqrt{N_u}}{2}\left(\begin{array}{c}
\delta_{\bm k,\bm q_\text{B}}+\delta_{\bm k,-\bm q_\text{B}}\\
\frac{1}{\imath}(\delta_{\bm k,\bm q_\text{B}}-\delta_{\bm k,-\bm q_\text{B}})\\0
\end{array}\right).
\end{equation}
Inserting this $\alpha$-independent vector into Eq.~(\ref{eq:ham_anisok}) yields
\begin{equation}
V_0=\frac{N_u}{4}\left(V^{xx}(\bm q_\text{B})+V^{yy}(\bm q_\text{B})\right).\label{eq:v0}
\end{equation}
Equivalent expressions $V_0=N_u(V^{xx}(\bm q_\text{B})+V^{zz}(\bm q_\text{B}))/4$ and $V_0=N_u(V^{yy}(\bm q_\text{B})+V^{zz}(\bm q_\text{B}))/4$ are obtained for spirals in the $x$-$z$ and $y$-$z$ planes, respectively.

\subsection{Helical spirals}
For helical spirals as realized in ZnCr$_2$Se$_4$, the normal direction of the spiral plane is parallel to the wave vector $\bm q_\text{B}$. Again, without loss of generality, we assume the spiral plane to be the $x$-$y$ plane such that $\bm q_\text{B}=(0,0,\pm\xi)$.

A natural and well-known mechanism that stabilizes helical spirals is the dipolar interaction between spins~\citeS{Utesov2018}. To show this, we consider the dipolar interaction
\begin{equation}
V=\frac{D}{2}\sum_{ij}\sum_{\alpha\beta}{}^{'}\,\frac{\bm S_{i\alpha}\cdot\bm S_{j\beta}-3(\bm S_{i\alpha}\cdot\hat{\bm d})(\bm S_{j\beta}\cdot\hat{\bm d})}{|\bm d|^3},\quad \bm d=\bm r_{i\alpha}-\bm r_{j\beta},\label{eq:dipolar}
\end{equation}
where $D>0$, the prime excludes $(i,\alpha)=(j,\beta)$, and lengths are measured in units of the cubic lattice constant $a$. This corresponds to $V^{\mu\nu}_{ij\alpha\beta}=D(\delta_{\mu\nu}-3\hat d_\mu\hat d_\nu)/|\bm d|^3$ in Eq.~(\ref{eq:ham_aniso_supp}). Two properties of $V^{\mu\nu}(\bm k)$ are important. First, the dipolar tensor is traceless for each $\bm d$, such that $\sum_\mu V^{\mu\mu}(\bm k)=0$ for all $\bm k$. Second, for $\bm q_\text{B}=(0,0,\xi)$ the mirror symmetry $x\leftrightarrow y$ yields $V^{xx}(\bm q_\text{B})=V^{yy}(\bm q_\text{B})$, while the twofold rotations about the cubic axes force all off-diagonal components $V^{\mu\nu}(\bm q_\text{B})$ to vanish. Together, this implies
\begin{equation}
V^{\mu\nu}(\bm q_\text{B})=V^{zz}(\bm q_\text{B})\,\text{diag}\left(-\frac{1}{2},-\frac{1}{2},1\right).
\end{equation}
According to Eq.~(\ref{eq:v0}), the energy difference between a helical spiral ($x$-$y$ plane) and a cycloidal spiral ($x$-$z$ or $y$-$z$ plane) is therefore
\begin{equation}
V_0^\text{helix}-V_0^\text{cycloid}=\frac{N_u}{4}\left(V^{yy}(\bm q_\text{B})-V^{zz}(\bm q_\text{B})\right)=-\frac{3N_u}{8}V^{zz}(\bm q_\text{B}),
\end{equation}
i.e., the dipolar interaction selects helical spirals whenever $V^{zz}(\bm q_\text{B})>0$.
This is the case for small $|\bm q_\text{B}|$, where the dipolar energy is dominated by its long-range part. Together with the tracelessness and cubic symmetry, it leads to $V^{\mu\nu}(\bm k)\propto\hat k_\mu\hat k_\nu-\frac{1}{3}\delta_{\mu\nu}$ with a positive prefactor for $\bm k\rightarrow0$, where $\hat{\bm k}=\bm k/|\bm k|$, and therefore to $V^{zz}(\bm q_\text{B})>0$ for $\bm q_\text{B}=(0,0,\xi)$.

Independent of the perturbation $V$ that selects helical spirals we can determine the masses of quartic modes in Eq.~(\ref{eq:quartic_mass}). Specifically, with $\bm q_\text{PG1}=(\xi,0,0)$ and $\bm q_\text{PG2}=(0,\xi,0)$, the cubic mirror symmetry that interchanges $x$ and $y$ yields $V^{xx}(\bm q_\text{B})=V^{yy}(\bm q_\text{B})$ and $V^{zz}(\bm q_{\text{PG1}})=V^{zz}(\bm q_{\text{PG2}})$. Similarly, the mirror symmetry interchanging $x$ and $z$ gives $V^{zz}(\bm q_{\text{PG1}})=V^{xx}(\bm q_\text{B})$. Inserting these properties into Eq.~(\ref{eq:quartic_mass}) immediately gives a vanishing mass matrix $M$. This means that for helical spirals and intact cubic symmetries no bilinear anisotropic spin interaction can give the quartic modes a non-zero mass in harmonic order.

\subsection{Cycloidal spirals}
Although not relevant for ZnCr$_2$Se$_4$, we briefly discuss a possible mechanism for the energetic selection of cycloidal spirals and investigate the mass generation of quartic modes in such spirals. 

We assume that $V$ is the most general symmetry-allowed anisotropic nearest-neighbor perturbation on the pyrochlore lattice in the {\it global} frame as given in Ref.~\citeS{Hanyan2017} in Eq.~(4). The couplings $J_{ij}^{\mu\nu}$ in this equation are given in Eq.~(6) and the site positions $\bm r_{i\alpha}$ for some reference unit cell $i=0$ are given in Eq.~(A1). Also note that the coupling matrices ${\bf J}_{01}$, ${\bf J}_{02}$, etc. in Eq.~(6) only refer to one tetrahedron type (e.g. up-tetrahedron), ${\bf J}_{01}\equiv{\bf J}_{01}^\text{up}$. Due to inversion symmetry with respect to lattice sites the corresponding matrices on the down tetrahedra are the same, ${\bf J}_{01}^\text{up}={\bf J}_{01}^\text{down}$. However, the bonds from sublattice 0 to sublattice 1 have the opposite orientation on the up- and down-tetrahedra.

Inserting all this information into Eqs.~(\ref{eq:v_matrix}) and (\ref{eq:subl_sum}) yields (setting the cubic lattice constant to $a=1$)
\begin{equation}
V^{xx}(\bm k)=8\mathcal{J}_2 \cos\left(\frac{k_y}{4}\right)\cos\left(\frac{k_z}{4}\right)+8\mathcal{J}_1 \cos\left(\frac{k_x}{4}\right)\cos\left(\frac{k_y}{4}\right)+8\mathcal{J}_1 \cos\left(\frac{k_x}{4}\right)\cos\left(\frac{k_z}{4}\right),\label{eq:vxx}
\end{equation}
while the other diagonal terms $V^{yy}(\bm k)$ and $V^{zz}(\bm k)$ follow by cyclic permutations of $x$, $y$ and $z$. Furthermore, the off-diagonal terms for $\mu\neq\nu$ are given by
\begin{equation}
V^{\mu\nu}(\bm k)=-8\mathcal{J}_3\sin\left(\frac{k_\mu}{4}\right)\sin\left(\frac{k_\nu}{4}\right).
\end{equation}
Here, $\mathcal{J}_1$, $\mathcal{J}_2$, $\mathcal{J}_3$ and $\mathcal{J}_4$ are the coupling constants denoted $J_1,\ldots,J_4$ in Ref.~\citeS{Hanyan2017} and explained below Eq.~(6); we use calligraphic letters to avoid confusion with the Heisenberg couplings of ZnCr$_2$Se$_4$. As a consequence of inversion symmetry and the sums over $\alpha$ and $\beta$ in Eq.~(\ref{eq:subl_sum}), the Dzyaloshinskii-Moriya interaction $\mathcal{J}_4$ drops out completely. 

For $\bm q_\text{B}=(0,0,\pm\xi)$ one finds
\begin{equation}\label{eq:v_j1j2}
V^{\mu\nu}(\bm q_\text{B})=8\text{ diag}\left(\mathcal{J}_1+(\mathcal{J}_1+\mathcal{J}_2)\cos\left(\frac{\xi}{4}\right),\mathcal{J}_1+(\mathcal{J}_1+\mathcal{J}_2)\cos\left(\frac{\xi}{4}\right),\mathcal{J}_2+2\mathcal{J}_1\cos\left(\frac{\xi}{4}\right)\right).
\end{equation}
One can easily check that for $\mathcal{J}_1>\mathcal{J}_2$, one has $V^{xx}(\bm q_\text{B})+V^{yy}(\bm q_\text{B})>V^{xx}(\bm q_\text{B})+V^{zz}(\bm q_\text{B})=V^{yy}(\bm q_\text{B})+V^{zz}(\bm q_\text{B})$ such that a helical spiral becomes energetically unfavorable and $V_0$ in Eq.~(\ref{eq:v0}) is minimized by a spiral in the $x$-$z$ or $y$-$z$ planes, corresponding to a cycloidal spiral. In fact, $V_0$ is the same for all spiral planes that contain the $z$ axis, and the orientation of the cycloidal plane is only selected by higher-order effects. For simplicity, we consider the $x$-$z$ and $y$-$z$ planes in the following.

Let us assume that the spiral is in the $x$-$z$ plane. According to Eq.~(\ref{eq:quartic_mass}), the mass matrix of quartic modes corresponding to this plane reads
\begin{equation}
M=-\frac{N_u}{4}\left(V^{xx}(\bm q_\text{B})+V^{zz}(\bm q_\text{B})\right)\text{diag}(1,1,1,1)
+\frac{N_u}{2}\,\text{diag}\left(V^{yy}(\bm q_{\text{PG1}}),V^{yy}(\bm q_{\text{PG1}}),V^{yy}(\bm q_{\text{PG2}}),V^{yy}(\bm q_{\text{PG2}})\right).\label{eq:quartic_mass_xz}
\end{equation}
Defining again $\bm q_\text{PG1}=(\xi,0,0)$ and $\bm q_\text{PG2}=(0,\xi,0)$ and using the symmetry that interchanges $y$ and $z$, one finds $V^{yy}(\bm q_{\text{PG2}})=V^{zz}(\bm q_\text{B})$. Furthermore, the $x\leftrightarrow y$ symmetry yields $V^{yy}(\bm q_{\text{PG1}})=V^{xx}(\bm q_{\text{PG2}})$ and in combination with the $y\leftrightarrow z$ symmetry one arrives at $V^{yy}(\bm q_{\text{PG1}})=V^{xx}(\bm q_{\text{B}})$. Inserting this into Eq.~(\ref{eq:quartic_mass_xz}) yields a non-vanishing result,
\begin{equation}
M=\frac{N_u}{4}\left(V^{xx}(\bm q_\text{B})-V^{zz}(\bm q_\text{B})\right)\text{diag}(1,1,-1,-1),
\end{equation}
which demonstrates that exactly one of the pseudo-Goldstone modes acquires a negative mass and therefore becomes unstable. More specifically, it follows from Eq.~(\ref{eq:v_j1j2}),
\begin{equation}
V^{xx}(\bm q_\text{B})-V^{zz}(\bm q_\text{B})=8(\mathcal{J}_1-\mathcal{J}_2)\left(1-\cos\left(\frac{\xi}{4}\right)\right),
\end{equation}
which is positive for $\mathcal{J}_1>\mathcal{J}_2$, such that the instability occurs at the quartic mode with wave vector $\bm q_\text{PG2}$ while the quartic mode with wave vector $\bm q_\text{PG1}$ acquires a positive mass. For a spiral in the $y$-$z$ plane one would come to the opposite conclusion. We note that, in contrast to the helical case, $V^{xx}(\bm q_\text{B})\neq V^{zz}(\bm q_\text{B})$ implies that the circular cycloid is not an exact stationary point of $H+V$; it acquires a slight elliptical deformation of order $V$ (involving higher harmonics of $\bm q_\text{B}$). This deformation modifies the masses of the quartic modes only at order $V^2$ and does not affect the above leading-order result. Overall, this demonstrates that in contrast to helical spirals, the quartic modes in cycloidal spirals acquire non-zero mass terms due to symmetry-allowed bilinear anisotropic spin interactions, where one mass is negative, triggering the formation of a double-$\bm q$ spiral.

\section{Symmetry-breaking Heisenberg perturbations}

We now consider the case of explicit symmetry breaking along the tetragonal [001] axis, for example due to elastic strain. Writing $H$ as a function of some strain parameter $\sigma$, we may treat this in the previously introduced formalism by defining
\begin{equation}
    H_0=H(0),\hspace{4em}V = H(\sigma)-H(0)
\end{equation}
Generically, this perturbation lifts the degeneracy between $\bm{q}_{\rm B}$ and $\bm{q}_{\mathrm{PG}n}$. If $H(\sigma)$ remains pure Heisenberg under strain, we have $V^{xx}=V^{yy}=V^{zz}=V(\bm{q})$. The condition for an unstable quartic mode then becomes
\begin{equation}
    M = \frac{N_u}{2}\left[V(\bm{q}_{\mathrm{PG}n}) - V(\bm{q}_{\mathrm B}) \right]  < 0~.
\end{equation}
Within the Luttinger-Tisza soluble space, this is the case if and only if $\bm{q}_{\rm B}$ is a metastable ordering vector. 

As a concrete example of such an instability, we consider the $J_1,J_2,J_3,J_4$
model studied in the main text, prepared by field cooling in strong [001] field.
The dipolar interaction selects helical order along [001] regardless of the
detailed energetics. \zncr\ is known to
undergo a structural transition at the ordering temperature, lowering the
symmetry $Fd\overline{3}m$ to the tetragonal
$I4_1/amd$~\citeS{zajdel_structure_2017_2}. We approximate this strain $\sigma$
in the [001] axis by the model  
\begin{align}
\bm{r}_i \mapsto \operatorname{diag}[1,1,(1-\sigma)]\bm{r}_{i,0}\,, \hspace{4em}
    J_{ij} \mapsto J_{ij}\left[1 - \alpha\left(\frac{\|\bm{r}_{i}-\bm{r}_{j}\|}{\|\bm{r}_{i,0}-\bm{r}_{j,0}\|}-1\right)\right]~.\label{eq:J_ij}
\end{align}
At leading order, we obtain $M = 7.21\times 10^{-3}\sigma\alpha|J_1|N_u$. 

When $\alpha\sigma > 0$, we find that ordering along [001] is preferred by a margin of approximately $0.0145\alpha\sigma|J_1|$, and $M>0$\,---\,the single-$\bm{q}$ order is stable, and there is no zero-temperature residue at $\bm{q}_{\rm PG}$.

When $\alpha\sigma < 0$, the situation is reversed. The lowest-energy ordering vectors are degenerate between [100] and [010], and indeed if one of these is chosen as $\bm{q}_{\rm B}$, then one of the remaining modes becomes a stable quadratic mode, while the other remains massless. However, if we prepare a metastable $\bm{q}_{\rm B}\,\|\, [001]$ order by e.g. field cooling, then an unstable quartic mode appears. As noted in the main text, this situation cannot arise from spontaneous magnetoelastic relaxation.

In the experiment, \zncr~\cite{zajdel_structure_2017} is known to distort with $\sigma>0$. Assuming shorter bonds to be stronger, the case $\alpha>0$, this renders the single-$\bm{q}$ order stable and so does not explain the observed multi-$\bm{q}$ order. The case $\alpha < 0$ is extremely unlikely in \zncr: it would require the magnetoelastic deformation to destabilize the very same ordering that induced it.

\section{Biquadratic perturbations to the Hamiltonian}
To investigate which symmetry-allowed perturbation can induce a multi-$\bm q$ deformation of a helimagnetic spiral, we next consider the effect of spin-isotropic biquadratic perturbations to the Hamiltonian, which have the general form
\begin{equation}
V=\frac{1}{2}\sum_{ij}\sum_{\alpha\beta}K_{ij\alpha\beta} ({\bm S}_{i\alpha}\cdot\bm S_{j\beta})^2.\label{eq:ham_biquad}
\end{equation}
Since $V$ is SO(3) symmetric, it cannot select a spiral plane, which means that $V_0$ is identical for all spiral planes.

In principle, we have to investigate Eq.~(\ref{eq:ham_biquad}) in the four-dimensional space of the soft modes parametrized in Eq.~(\ref{eq:parametrize_soft}). However, due to translation invariance, $M$ must again be diagonal and, without loss of generality, it suffices to only consider the deformation with amplitude $a_{\bm q_\text{PG1}}$ in Eq.~(\ref{eq:parametrize_soft}) yielding
\begin{equation}
\Delta_{i\alpha}^\perp=a_{\bm q_\text{PG1}} \cos({\bm q}_\text{PG1}\cdot{\bm r}_{i\alpha}).\label{eq:parametrize_biquad}
\end{equation}
In the following, we calculate the mass term corresponding to this deformation, which appears in quadratic order $\sim\!a^2_{\bm q_\text{PG1}}$ in an expansion of $V$. The other soft modes contained in Eq.~(\ref{eq:parametrize_soft}) yield identical results, since the modes $a_{\bm q_\text{PG1}}$ and $b_{\bm q_\text{PG1}}$ are related by a translation along $\bm q_\text{PG1}$, and the modes at $\bm q_\text{PG1}$ and $\bm q_\text{PG2}$ are related by the mirror symmetry $x\leftrightarrow y$. Inserting Eq.~(\ref{eq:parametrize_biquad}) into Eq.~(\ref{eq:perturbation2}) yields, up to quadratic order, the following spin configuration
\begin{equation}
\bm S_{i\alpha}=\left(
\begin{array}{c}
\cos(\bm q_\text{B}\cdot \bm r_{i\alpha})-\frac{1}{2}a_{\bm q_\text{PG1}}^2\cos(\bm q_\text{B}\cdot \bm r_{i\alpha})\cos^2(\bm q_\text{PG1}\cdot \bm r_{i\alpha})\\
\sin(\bm q_\text{B}\cdot \bm r_{i\alpha})-\frac{1}{2}a_{\bm q_\text{PG1}}^2\sin(\bm q_\text{B}\cdot \bm r_{i\alpha})\cos^2(\bm q_\text{PG1}\cdot \bm r_{i\alpha})\\
a_{\bm q_\text{PG1}} \cos(\bm q_\text{PG1}\cdot \bm r_{i\alpha})
\end{array}
\right).\label{eq:biquad_perturb}
\end{equation}
For a pair of sites $(i\alpha,j\beta)$ with $\bm d=\bm r_{j\beta}-\bm r_{i\alpha}$ and $g_{i\alpha}\equiv\cos(\bm q_\text{PG1}\cdot\bm r_{i\alpha})$, Eq.~(\ref{eq:biquad_perturb}) gives, up to second order in $a_{\bm q_\text{PG1}}$,
\begin{equation}
\bm S_{i\alpha}\cdot\bm S_{j\beta}=\cos(\bm q_\text{B}\cdot\bm d)\left[1-\frac{1}{2}a_{\bm q_\text{PG1}}^2\left(g_{i\alpha}^2+g_{j\beta}^2\right)\right]+a_{\bm q_\text{PG1}}^2g_{i\alpha}g_{j\beta},
\end{equation}
and therefore
\begin{equation}
(\bm S_{i\alpha}\cdot\bm S_{j\beta})^2=\cos^2(\bm q_\text{B}\cdot\bm d)+a_{\bm q_\text{PG1}}^2\cos(\bm q_\text{B}\cdot\bm d)\left[2g_{i\alpha}g_{j\beta}-\cos(\bm q_\text{B}\cdot\bm d)\left(g_{i\alpha}^2+g_{j\beta}^2\right)\right]+\mathcal{O}(a_{\bm q_\text{PG1}}^4).
\end{equation}
When summing over all unit cells $i$ in Eq.~(\ref{eq:ham_biquad}) at fixed sublattices and fixed bond vector $\bm d$, the products of the $g_{i\alpha}$ can be replaced by their lattice averages, $g_{i\alpha}g_{j\beta}\rightarrow\frac{1}{2}\cos(\bm q_\text{PG1}\cdot\bm d)$ and $g_{i\alpha}^2,g_{j\beta}^2\rightarrow\frac{1}{2}$, because the remaining terms oscillate with wave vector $2\bm q_\text{PG1}$ and average to zero. This yields the general result
\begin{equation}
V=\chi'+\frac{a_{\bm q_\text{PG1}}^2}{2}\sum_{ij}\sum_{\alpha\beta}K_{ij\alpha\beta}\cos(\bm q_\text{B}\cdot\bm d)\left[\cos(\bm q_\text{PG1}\cdot\bm d)-\cos(\bm q_\text{B}\cdot\bm d)\right]+\mathcal{O}(a_{\bm q_\text{PG1}}^4),\label{eq:biquad_general}
\end{equation}
where $\chi'$ is the energy of the undeformed spiral.
 
In the simplest case, $K_{ij\alpha\beta}$ is a non-zero constant $\kappa$ only on nearest-neighbor pyrochlore bonds. Each primitive unit cell contains 12 such bonds, with bond vectors (in units of the cubic lattice constant) $\bm d=(\pm\frac14,\pm\frac14,0)$, $(\pm\frac14,0,\pm\frac14)$, and $(0,\pm\frac14,\pm\frac14)$, i.e., four bonds in each of the three cubic planes. With $\bm q_\text{B}=(0,0,\xi)$, $\bm q_\text{PG1}=(\xi,0,0)$, and the abbreviation $u\equiv\cos(\xi/4)$, the bracket in Eq.~(\ref{eq:biquad_general}) multiplied by $\cos(\bm q_\text{B}\cdot\bm d)$ takes the values $u-1$ for bonds in the $x$-$y$ plane, $u(u-u)=0$ for bonds in the $x$-$z$ plane, and $u(1-u)$ for bonds in the $y$-$z$ plane. Since the sum over $i,j$ counts each bond twice, we obtain
\begin{equation}
V=\chi'+4N_u\kappa\,a_{\bm q_\text{PG1}}^2\left[(u-1)+u(1-u)\right]=\chi'-4N_u\kappa\,a_{\bm q_\text{PG1}}^2(1-u)^2=\chi'-16N_u\kappa\,a_{\bm q_\text{PG1}}^2\sin^4\left(\frac{\xi}{8}\right).
\end{equation}
This shows that the quartic modes indeed acquire a negative mass for $\kappa>0$ which can explain the observed elastic signal at the pseudo-Goldstone wave vector of ZnCr$_2$Se$_4$.

\section{Supplemental Numerical Results}

To exemplify the general principle uncovered in the main text, we consider the fourth-neighbor pyrochlore Heisenberg model introduced by Ref.~\citeS{PG2017PRX}, with an additional biquadratic interaction
\begin{equation}
    H_{\rm Heis}=\sum_{n=1}^4\sum_{\langle ij\rangle_n} J_n \bm{S}_i\cdot \bm{S}_j + \sum_{\langle ij \rangle_1} \frac{\kappa}{S(S+1)} (\bm{S}_i\cdot \bm{S}_j)^2\,,
    \label{eq:h_full}
\end{equation}
where $\langle ij \rangle_n$ denotes the $n$th nearest neighbor by geometric distance. For simplicity, we take the two inequivalent third-neighbor exchange couplings $J_{3a},J_{3b}$ (one of which has a magnetic site in the center) to be of equal strength.

Despite the presence of biquadratic interactions, one may substitute the spiral Ansatz of \eqref{eq:planar_spiral} into \eqref{eq:h_full} and re-minimize the variational energy in $\bm{q}$. At least on the cubic axis, this can be done analytically:
\begin{equation}
 (8J_3 + 4J_2+S(S+1)\kappa) \cos(2\xi a) +
    J_4 (12 \cos^2(2\xi a) -2) +  2 J_2 = -J_1,
\end{equation}
As claimed in the main text, we see that $\xi$'s dependence on $J_3$ and $\kappa$ can be expressed as a function of $(8J_3 +S(S+1)\kappa)$.

\begin{figure}
    \includegraphics{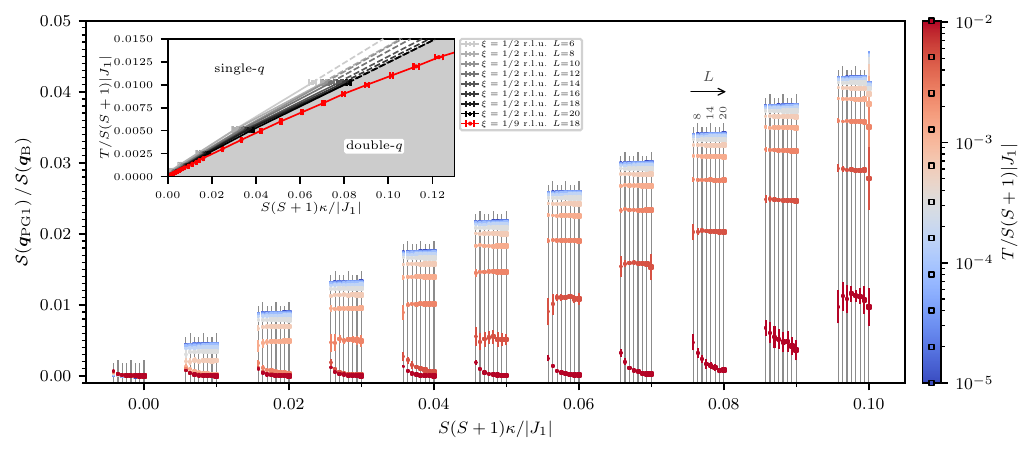}
    \caption{Scaling of the pseudo-Goldstone peak height $\cmcal{S}(\bm{q}_{\rm PG1})$ with system size, cf. \autoref{fig:numerics} of the main text. Each column shows increasing $L$ from left to right, at evenly spaced $L=6,8,\dots,20$. Each system size $L$ is manually offset in the $x$-axis for legibility.  Inset: $\kappa-T$ phase diagram in the $\xi=1/2\;\text{r.l.u.}$ plane, obtained by extrapolating the $x$-intercept of each $\cmcal{S}(q)- \kappa$ trace.}\label{fig:scaling_phasedia}
\end{figure}

In order to exclude the possibility of finite-size effects in the simulations, we studied a $J_1,J_2,J_3,J_4,\kappa$ model over a range of eight system sizes. In order to achieve this within a reasonable computational timeframe, we retuned the model presented in the main text to have a commensurate $\xi =1/2\;\text{r.l.u.}$ instead of $4/9\;\text{r.l.u.}$ This difference amounts to using $(J_3+S(S+1)\kappa/8)=0.176604|J_1|$, only $\sim$7\% greater than that used in the main text. 

From \autoref{fig:scaling_phasedia}, excellent convergence in system size can be seen at all $T\le 0.005|J_1|$. Firstly, this confirms that the pseudo-Goldstone peak scales with system size at the same rate as the Bragg peak; a necessary feature of a genuine double-$\bm q$ order. Secondly, we can be confident that any finite-size effects lingering in the $L=18$ simulation used for the main text are small. 

For each single-temperature series, we fitted all data above the cutoff $\cmcal{S}(\bm{q}_{\rm PG1})/\cmcal{S}(\bm{q}_{\rm B})=0.001$ with a straight line, then extrapolated back to the point $\cmcal{S}(q_{\rm PG1})=0$. The resulting critical $\kappa$'s are plotted in the inset of \autoref{fig:scaling_phasedia}, showing a clear linear relationship between $\kappa_c$ and $T_c$. This suggests a straightforward competition of energy scales between $\kappa$ and the entropic cost of condensing the $\mathbb{Z}_2$ order parameter.

\subsection{Estimation of $\kappa$ from experimental data.}

Remarkably, the observed ratio of Bragg peaks in the experiments on \zncr\ allows us to give a crude estimate for $\kappa$.  
Looking at \autoref{fig:experiment}, we assume that the low-temperature ZFC peak is one of three symmetry-equivalent spiral Bragg peaks, while the 0~T FC peak arises from an equal population of two double-$\bm{q}$ spiral orders. This gives $\cmcal{S}(\bm{q}_{\rm PG1}) / \cmcal{S}(\bm{q}_{\rm B})=(2\cmcal{S}_{\rm FC})/(3\cmcal{S}_{\rm ZFC})\simeq8\cdot10^{-3}$. Taking $J_1=-2.876\,\rm{meV}$~\citeS{PG2017PRX}, we see that the lowest experimental temperature ${T}=1.5\,\mathrm{K}=0.0120\,{S(S+1)|J_1|/k_B}$. Both of these are plotted as red dashed lines in \autoref{fig:numerics} of the main text, from which one may read off $\kappa S(S+1) \simeq 0.13 |J_1|$. 

This result should be interpreted with some care\,---\,the position of the double-$\bm{q}$ $T_c$ depends sensitively on material details that cannot be constrained by elastic scattering alone. A more precise determination may be possible by combining several data sources, a worthwhile endeavor which we leave to future work.

\makeatletter
\setcounter{NAT@ctr}{0}   %
\makeatother

\bibliographystyleS{apsrev4-2}
\bibliographyS{bibliography_supp}

\end{document}